\documentclass[aps,prl,twocolumn,floatfix,showpacs]{revtex4}
\usepackage{graphics}
\usepackage{graphicx}
\usepackage{amssymb}
\usepackage{amsfonts}
\usepackage{amsmath}
\usepackage{bm}
\usepackage{epsf}
\usepackage{version}

\newcommand \be  {\begin{equation}}
\newcommand \ee  {\end{equation}}
\newcommand \bea {\begin{eqnarray} }
\newcommand \eea {\end{eqnarray}}

\newcommand \bd  {\begin{details}}
\newcommand \ed  {\end{details}}
\begin{document}

\title{Kramers degeneracy in a magnetic field \\ 
and Zeeman spin-orbit coupling in antiferromagnetic 
conductors}
\author{Revaz Ramazashvili}
\thanks{Present address: Department of Physics 
and Astronomy, University of South Carolina, 
Columbia, SC 29208, USA.}
\affiliation{
ENS, LPTMS, UMR8626, B\^at. 100, 
Universit\'e Paris-Sud, 91405 Orsay, France
}

\date{\today}

% \includeversion{details}
\excludeversion{details}

% \Large

\begin{abstract}
In this article, I analyze the symmetries 
and degeneracies of electron eigenstates 
in a commensurate collinear antiferromagnet. 
In a magnetic field transverse to the 
staggered magnetization, a hidden anti-unitary 
symmetry protects double degeneracy of the Bloch  
eigenstates at a special set of momenta. 
In addition to this `Kramers degeneracy' subset, 
% the degeneracy manifold in a transverse field 
the manifold of momenta, labeling the doubly 
degenerate Bloch states in the Brillouin zone, 
may also contain an `accidental de\-ge\-ne\-ra\-cy' 
subset, that is not protected by symmetry and 
that may change its shape under perturbation. 
These degeneracies give rise to a substantial 
momentum dependence of the transverse $g$-factor 
in the Zeeman coupling, turning the latter into 
a spin-orbit interaction. 

I discuss a number of materials, where Zeeman 
spin-orbit coupling is likely to be present, 
and outline the simplest properties and 
experimental consequences of this interaction, 
that may be relevant to systems from chromium 
to borocarbides, cuprates, hexaborides, iron 
pnictides, as well as organic and heavy fermion 
conductors. 
\end{abstract}


\maketitle

% \bigskip

\section{I. Introduction}
 
Antiferromagnetism is widespread in materials with 
interesting electron properties. Chromium \cite{fawcett_1} 
and its alloys \cite{fawcett_2,kulikov}, numerous 
borocarbides \cite{mueller}, electron- and hole-doped 
cuprates \cite{birgeneau,tranquada}, iron pnictides  
\cite{cruz}, various organic \cite{chaikin} and heavy 
fermion \cite{kuramoto,robinson,flouquet} compounds 
all have an antiferromagnetic state present in their 
phase diagram. The physics of these antiferromagnetic 
phases has been a subject of active research. 

In this article, I study the response of electron 
Bloch eigenstates in an antiferromagnet to a weak 
magnetic field. 
I concentrate on the simplest case: a centrosymmetric 
doubly commensurate collinear antiferromagnet, shown 
schematically in Fig. \ref{fig:AF_real_space}, where 
the magnetization density at any point in space 
is pa\-ral\-lel or anti-pa\-ral\-lel to a single 
fixed direction ${\bf n}$ of the staggered magnetization, 
and changes sign upon primitive translation of the 
underlying lattice. 

\begin{details}
Below, I show that, in a magnetic field transverse 
to the staggered magnetization, a hidden anti-unitary 
symmetry protects the Kramers degeneracy of Bloch 
eigenstates at a special set of momenta. This 
degeneracy gives rise to a peculiar spin-orbit 
coupling, whose emergence and basic properties, 
along with the degeneracy itself, are the main 
results of this work.
\end{details}

In a paramagnet, the double degeneracy of the Bloch 
eigenstates is commonly attributed to symmetry under 
time reversal $\theta$ -- and, indeed, perturbations 
that break time reversal symmetry (such as ferromagnetism 
or a magnetic field) do tend to remove the degeneracy. 
Yet violation of $\theta$ alone does not preclude 
degeneracy: in a commensurate centrosymmetric 
N\'eel antiferromagnet, as in a paramagnet, 
all Bloch eigenstates enjoy a Kramers degeneracy 
\cite{herring1} in spite of time reversal symmetry 
being broken in the former, but not in the latter.

In an antiferromagnet, the staggered magnetization sets 
a special direction ${\bf n}$ in electron spin space, 
ma\-king it anisotropic. A magnetic field along ${\bf n}$ 
removes the de\-ge\-ne\-ra\-cy of all Bloch eigenstates, 
as it does in a pa\-ra\-mag\-net. By contrast, in a 
transverse field, a hidden anti-unitary symmetry 
protects the Kramers degeneracy of Bloch 
eigenstates at a special set of momenta. 

Generally, in $d$ dimensions, the manifold 
of momenta, corresponding to doubly degenerate 
Bloch states in a transverse field is 
$(d-1)$-dimensional; within a subset 
of this manifold, the degeneracy is dictated 
by sym\-met\-ry. This is in marked contrast 
with what happens in a paramagnet, where 
an arbitrary magnetic field lifts the 
degeneracy of \textit{all} Bloch eigenstates. 
For brevity, in this article I often refer 
to the manifold of momenta, labeling the 
degenerate Bloch states in the Brillouin zone, 
as to the `degeneracy manifold'.

As a consequence of the Kramers degeneracy of the 
special Bloch states in a transverse field, the 
transverse component $g_\perp$ of the electron 
$g$-tensor vanishes for such states. Not being 
identically equal to zero, $g_\perp$ must, therefore, 
carry a substantial momentum dependence, and the 
Zeeman coupling $\mathcal{H}_{ZSO}$ must take the 
form  
\be 
\label{eq:ZSO}
\mathcal{H}_{ZSO}
 = 
 - \mu_B 
\left[ 
g_\| 
({\bf H_\| \cdot \bm{\sigma}}) 
 +
g_\perp ({\bf p}) 
({\bf H_\perp \cdot \bm{\sigma}})
\right], 
\ee
where 
$
{\bf H}_\| 
 =
({\bf H} \cdot {\bf n}) 
{\bf n}
$  
and
$
{\bf H}_\perp
 =
 {\bf H}
 - 
{\bf H}_\| 
$ 
are the longitudinal and transverse components 
of the magnetic field with respect to the unit 
vector ${\bf n}$ of the staggered magnetization, 
$
\mu_B 
% \equiv \frac{e \hbar}{2 m_0 c}
$ 
is the Bohr magneton, while $g_\|$ and 
$g_\perp ({\bf p})$ are the longitudinal 
and transverse components of the $g$-tensor.  

This significant momentum dependence of 
$g_\perp ({\bf p})$ turns the common Zeeman 
coupling into a kind of spin-orbit interaction 
$\mathcal{H}_{ZSO}$  (\ref{eq:ZSO}), whose 
appearance and key pro\-per\-ties are the focus 
of this work. Zeeman spin-orbit coupling may 
manifest itself spectacularly in a number 
of ways, which will be mentioned below and 
discussed in detail elsewhere.

The symmetry properties of wave functions in magnetic 
crystals have been studied by Dimmock and Wheeler  
\cite{dimmock2}, who pointed out, among other things, 
that magnetism not only lifts degeneracies by obviously  
lowering the symmetry, but also may introduce new ones. 
This may happen at the magnetic Brillouin zone (MBZ) 
boundary, under the necessary condition that the 
magnetic unit cell be larger than the paramagnetic 
one \cite{dimmock2}. 

For a N\'eel antiferromagnet on a lattice of square 
symmetry, the response of the electron states to 
a magnetic field was studied in Ref. \cite{braluk} 
using symmetry arguments, and in Ref. \cite{bralura} 
within a weak coupling model. The present work 
is a detailed presentation of recent results 
\cite{rr_sym}. It revisits Ref. \cite{braluk}, 
extends it to an arbitrary crystal symmetry and 
to a finite as opposed to infinitesimal magnetic 
field, and uncovers a rich interplay between the 
symmetry of magnetic structure and that of the 
underlying crystal lattice. At the same time, 
the present work extends Ref. \cite{dimmock2} 
by allowing for an external magnetic field -- 
to show how, at special momenta, the Kramers 
degeneracy in an antiferromagnet may persist 
even in a transverse magnetic field.

This work treats antiferromagnetic order as static, 
neglecting both its classical and quantum fluctuations. 
This excludes from consideration strongly fluctuating  
antiferromagnetic states such as those near a continuous 
phase transition, be it a finite-temperature N\'eel 
transition or a quantum ($T = 0$) critical point. At 
the same time, the single-electron Bloch eigenstates, 
considered hereafter, must be well-defined. As in a 
normal Fermi liquid state, this does not rule out 
strong interaction between electrons, but simply 
requires temperatures well below the Fermi energy. 
Finally, to justify the neglect of quantum 
fluctuations, the ordered magnetic moment must be 
of the order of or greater than the Bohr magneton. 

As a consequence, the present theory applies 
to antiferromagnets \textit{(i)} deep inside 
a commensurate long-range antiferromagnetic 
state, and far enough from any con\-ti\-nu\-ous 
N\'eel transition, finite-temperature or quantum,  
\textit{(ii)} with an ordered moment noticeable 
on the scale of the Bohr magneton, and 
\textit{(iii)} far below both the N\'eel and the 
effective Fermi temperatures. All materials 
mentioned in Section IV are meant to be 
considered under these conditions.

The article is organized as follows. 
Section II opens with a reminder of how, in spite of 
broken time reversal symmetry, all Bloch eigenstates 
in a commensurate collinear antiferromagnet retain 
Kramers de\-ge\-ne\-ra\-cy, provided there is 
an inversion center \cite{herring1}. 
Then I show how, even in a transverse magnetic field, 
a hidden symmetry of antiferromagnetic order may protect 
the Kramers degeneracy for certain Bloch states.

Section III establishes several properties of the 
single-electron spectrum in a weakly coupled 
antiferromagnet, subject to a transverse magnetic field. 

Section IV contains the analysis of simple examples, 
that may be relevant to specific materials from 
chromium to organic conductors, from borocarbides 
to underdoped cuprates and to various heavy fermion 
metals. The Discussion section reviews the findings, 
and examines them in the light of earlier work, 
while the Appendices present various technical 
details. 
% Appendix A derives an identity, used in the main 
% text to demonstrate the Kramers degeneracy, 
% Appendix B describes how the sublattice canting 
% affects the degeneracy manifold in a transverse 
% field, Appendix C derives the dimensionality 
% of the manifold of degenerate Bloch states 
% by a symmetry argument, and Appendix D 
% examines several details of Ref. \cite{braluk}.

\section{II. General arguments}  

It is convenient to begin by describing the crystal 
host symmetry in the absence of magnetism, with 
the average magnetization density notionally set 
to zero \cite{dimmock2}. I refer to this as to 
the paramagnetic state symmetry, even though the 
symmetry of the actual paramagnetic state may be 
different, for instance due to a lattice distortion upon 
transition. Unitary symmetries of the paramagnetic state 
form a group, $h$, which includes a set of elementary 
translations ${\bf T_a}$ by primitive translation vectors  
$\bf{a}$. Time reversal $\theta$ being indeed a symmetry 
of the paramagnetic state, the full symmetry group $g$ 
of the paramagnetic state includes $h$, and products 
of $\theta$ with each element of $h$: 
$g = h + \theta \cdot h$; put otherwise, $h$ 
is an invariant unitary subgroup of $g$. 

Antiferromagnetic order couples to the 
electron spin $\bm{\sigma}$ via the exchange term 
$({\bf \Delta_r} \cdot \bm{\sigma})$, where 
${\bf \Delta_r}$ is proportional to the average 
microscopic magnetization at point ${\bf r}$. 
In keeping with the arguments of the Introduction, 
fluctuations of ${\bf \Delta_r}$ are neglected. 
Being of relativistic origin, spin-orbit couplings  
of the crystal lattice to the electron spin and to 
the magnetization density are also neglected. This 
is a good approximation in a broad range of problems, 
at the very least at temperatures above the scale 
set by the spin-orbit coupling (see Section V for 
details). In this `exchange symmetry' approximation  
\cite{andreev}, magnetization density and electron 
spin are assigned to a separate space, independent 
of the real space of the crystal; this makes 
coordinate rotations and other point symmetries 
inert with respect to ${\bf \Delta_r}$ and 
$\bm{\sigma}$. 

\begin{figure}[h]
 \hspace{3cm}
 \epsfxsize=8cm
 \epsfbox{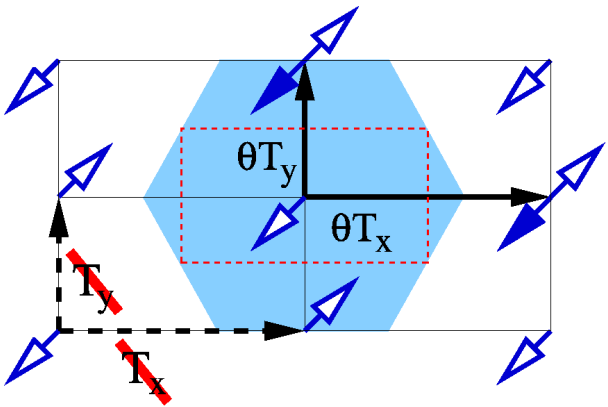}
 \vspace{15pt}
\caption{(color online). 
Doubly commensurate collinear antiferromagnet 
on a simple rectangular lattice. 
In the absence of magnetism, time reversal $\theta$ 
and primitive translations ${\bf T}_x$ and ${\bf T}_y$, 
shown by dashed arrows, are symmetry operations. In the  
antiferromagnetic state, neither of the three remains 
a symmetry, but the products $\theta {\bf T}_x$ and 
$\theta {\bf T}_y$, shown by solid arrows, do, 
as illustrated by filled spin arrows. Small dashed 
rectangle at the center is the Wigner-Seitz cell boundary 
in the paramagnetic state, while the shaded hexagon is its  
antiferromagnetic counterpart. Notice that neither 
of the point group operations interchanges the two  
sublattices, hence any point symmetry of the lattice,  
including inversion $\mathcal{I}$, remains a symmetry 
of the antiferromagnetic state. 
} 
\label{fig:AF_real_space}
\end{figure}

A nonzero ${\bf \Delta_r}$ changes sign under time 
reversal $\theta$, and removes the symmetry under 
primitive translations ${\bf T_a}$, thus reducing 
the symmetry with respect to that of paramagnetic state. 
In a doubly commensurate collinear antiferromagnet, 
${\bf \Delta_r}$ changes sign upon ${\bf T_a}$: 
${\bf \Delta_{r+a} = - \Delta_r}$, while 
${\bf T}_{\bf a}^2$ leaves ${\bf \Delta_r}$ intact: 
${\bf \Delta}_{{\bf r}+2{\bf a}} = {\bf \Delta_r}$. 
Even though neither $\theta$ nor ${\bf T_a}$ 
remain a symmetry, their product $\theta {\bf T_a}$ 
does (see Fig. \ref{fig:AF_real_space}).
In a system with inversion center, 
so does $\theta {\bf T_a} \mathcal{I}$, 
where $\mathcal{I}$ is inversion. The importance of 
the combined symmetry $\theta {\bf T_a} \mathcal{I}$ 
will become clear in the next subsection. 

%%%%%%%%%% Figure %%%%%%%%%% 
%%%%%%%%%%%%%%%%%%%%%%%%%%%

Together with the uniaxial character 
(${\bf \Delta_r}$ at any point ${\bf r}$ pointing 
along or against the single direction ${\bf n}$ 
of staggered magnetization), these relations define  
a commensurate collinear N\'eel antiferromagnet via 
transformation properties of its microscopic 
magnetization density. 

\subsection{Kramers degeneracy in zero field}

The combined  anti-unitary symmetry 
$\theta {\bf T_a} \mathcal{I}$ gives rise 
to a Kramers degeneracy \cite{herring1}: 
If $| {\bf p} \rangle$ is a Bloch eigenstate 
at momentum ${\bf p}$, then 
$\theta {\bf T_a} \mathcal{I} | {\bf p} \rangle$ 
is degenerate with $| {\bf p} \rangle$. 
Since $\theta$ and $\mathcal{I}$ both invert 
the momentum, both $| {\bf p} \rangle$ and 
$\theta {\bf T_a} \mathcal{I} | {\bf p} \rangle$ carry the 
same momentum label ${\bf p}$. Formally, this is verified 
by the action of any translation ${\bf T}_{\bf b}$, that 
remains a symmetry of the antiferromagnetic state:
\bea 
 & &  
\nonumber
{\bf T}_{\bf b}
\theta {\bf T_a} \mathcal{I} | {\bf p} \rangle 
 = 
\theta {\bf T_a} 
{\bf T}_{\bf b}  
\mathcal{I} | {\bf p} \rangle 
 = 
\theta {\bf T_a} \mathcal{I} 
{\bf T}_{- \bf b} 
| {\bf p} \rangle 
 = \\
 & = & 
\theta {\bf T_a} \mathcal{I} 
e^{-i \bf p \cdot b} 
| {\bf p} \rangle
 = 
e^{i \bf p \cdot b} 
\theta {\bf T_a} \mathcal{I} 
| {\bf p} \rangle. 
\eea 
At the same time, $| {\bf p} \rangle$ and 
$\theta {\bf T_a} \mathcal{I} | {\bf p} \rangle$ 
are orthogonal. 
This follows from Eqn. (\ref{eq:antiproduct2}) of Appendix A  
as soon as one chooses   
$\mathcal{O} = {\bf T_a} \mathcal{I}$, 
$| \psi \rangle = | {\bf p} \rangle$, and  
$| \phi \rangle = {\bf T_a} \mathcal{I} \theta | {\bf p} \rangle$.   
Recalling that $({\bf T_a} \mathcal{I})^2 = - \theta^2 =1$, 
and hence $(\theta {\bf T_a} \mathcal{I})^2 =-1$, one finds 
\be
\label{eq:Kramers}
\langle \bf{p} | \mathcal{I} {\bf T_a} \theta | \bf{p} \rangle 
 = - 
\langle \bf{p} |  \mathcal{I} {\bf T_a} \theta | \bf{p} \rangle. 
\ee
Thus, in spite of broken time reversal symmetry, 
in a centrosymmetric commensurate N\'eel antiferromagnet 
\textit{all} Bloch states retain a Kramers degeneracy.

\subsection{Kramers degeneracy in a transverse field}

Generally, a magnetic field ${\bf H}$ lifts this 
degeneracy. However, in a transverse field, 
a hidden anti-unitary symmetry may protect the 
degeneracy at a special set of points in the 
Brillouin zone, as I show below. 

In an antiferromagnet, subject to a magnetic 
field ${\bf H}$, the single-electron Hamiltonian 
takes the form  
\be
\label{eq:Hamiltonian_generic}
\mathcal{H} 
 = 
\mathcal{H}_0 + 
({\bf \Delta_r} \cdot \bm{\sigma}) 
 - ({\bf H} \cdot \bm{\sigma}), 
\ee
where the `paramagnetic' part $\mathcal{H}_0$ is 
invariant under independent action of ${\bf T_a}$ 
and $\theta$, and $g \mu_B$ is set to unity. 
In the absence of the field, all Bloch eigenstates 
of Hamiltonian (\ref{eq:Hamiltonian_generic}) are 
doubly degenerate by virtue of Eqn. (\ref{eq:Kramers}).

%%%%%%%%%% Figure %%%%%%%%
%%%%%%%%%%%%%%%%%%%%%%%%%%
\begin{figure}[h]
 \hspace{0cm}
 \epsfxsize=5cm
 \epsfbox{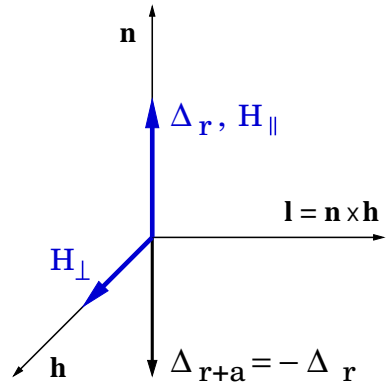}
 \vspace{15pt}
\caption{(color online). 
Relative orientation of ${\bf \Delta_r}$, 
${\bf \Delta_{r+a}}$, ${\bf H}_\|$ and 
${\bf H}_\perp$. To see the combined 
symmetries of Table \ref{table:symmetries},  
notice that $\theta$ flips both ${\bf \Delta_r}$ 
and ${\bf H}$, while ${\bf T_a}$ leaves ${\bf H}$ 
intact, but inverts ${\bf \Delta_r}$. 
Unit vectors ${\bf n}$, ${\bf h}$ and 
${\bf l = n \times h}$ are defined as shown.
} 
\label{fig:dhl}
\end{figure}
%%%%%%%%%%%%%%%%%%%%%%%%%%%
%%%%%%%%%%%%%%%%%%%%%%%%%%%
Consider the symmetries of Hamiltonian 
(\ref{eq:Hamiltonian_generic}), involving a combination 
of an elementary translation ${\bf T_a}$, time reversal 
$\theta$, or a spin rotation ${\bf U_m}(\phi)$ around  
an axis ${\bf m}$ by an angle $\phi$. These symmetries 
are listed in Table \ref{table:symmetries}; the relative  
orientation of ${\bf \Delta_r}$, ${\bf H}_\|$ and 
${\bf H}_\perp$ is shown in Fig. \ref{fig:dhl}. 
\begin{table}
\begin{center}
\caption{
Symmetries of a collinear doubly commensurate 
antiferromagnet. The left column is for zero field, 
central column for a transverse magnetic field 
${\bf H}_\perp$, and the right column for a 
longitudinal field ${\bf H}_\|$. 
As above, ${\bf T_a}$ denotes elementary translation 
by ${\bf a}$. As shown in Fig. \ref{fig:dhl}, unit 
vector ${\bf n}$ is collinear with ${\bf \Delta_r}$, 
unit vector ${\bf h}$ points along ${\bf H}_\perp$, 
and unit vector ${\bf l}$ is defined via 
${\bf l} = {\bf n} \times {\bf h}$. 
For a general orientation of the field 
(${\bf H}_\| \neq 0$, ${\bf H}_\perp \neq 0$), 
not shown in this table, 
${\bf U_l}(\pi) \theta$ is the only surviving symmetry, 
where ${\bf U_m}(\phi)$ denotes spin rotation 
by angle $\phi$ around the ${\bf m}$ axis. 
}
\vspace{15pt}
\begin{tabular}{c|c|c}
\hline
\hline
$({\bf \Delta_r} \cdot \bm{\sigma})$ 
 & 
$({\bf \Delta_r} \cdot \bm{\sigma})
 + ({\bf H}_\perp \cdot \bm{\sigma})$ 
 &
$({\bf \Delta_r} \cdot \bm{\sigma})
 + ({\bf H}_\| \cdot \bm{\sigma})$  \\
 & & \\ 
\hline
 ${\bf U_n}(\phi)$  & -- & ${\bf U_n}(\phi)$ \\
${\bf U_h}(\pi) {\bf T_a}$ & 
${\bf U_h}(\pi) {\bf T_a}$ & -- \\
${\bf U_l}(\pi) {\bf T_a}$ &  -- & -- \\
${\bf T_a} \theta$ & 
${\bf U_n}(\pi) {\bf T_a} \theta$ & -- \\
${\bf U_h}(\pi) \theta$ & -- & ${\bf U_h}(\pi) \theta$ \\
${\bf U_l}(\pi) \theta$ & ${\bf U_l}(\pi) \theta$ &  
${\bf U_l}(\pi) \theta$ \\
\hline
\hline
\end{tabular}
\label{table:symmetries}
\end{center}
\end{table}
The transverse field ${\bf H}_\perp$ breaks the 
symmetries ${\bf U_n}(\phi)$ and ${\bf T_a} \theta$,  
% (both affect ${\bf H}_\perp$), 
but preserves their combination at $\phi = \pi$, 
i.e. ${\bf U_n}(\pi) {\bf T_a} \theta$. 
 Acting on an exact Bloch state 
$| {\bf p} \rangle$ at momentum ${\bf p}$, 
this combined anti-unitary ope\-ra\-tor 
creates a degenerate partner eigenstate 
${\bf U_n}(\pi) {\bf T_a} \theta | {\bf p} \rangle$, 
which is orthogonal to $| {\bf p} \rangle$ everywhere in the 
Brillouin zone, unless $({\bf p \cdot a})$ is an integer 
multiple of $\pi$ (in other words, unless ${\bf p}$ 
lies at a paramagnetic Brillouin zone boundary): 
\be 
\label{eq:thetaTU}
\langle 
{\bf p}
 | 
{\bf U_n}(\pi) {\bf T_a} \theta 
 | {\bf p}  
\rangle 
 = 
e^{-2 i ({\bf p \cdot a})}
\langle 
{\bf p}
 | 
{\bf U_n}(\pi) {\bf T_a} \theta
 | {\bf p}  
\rangle. 
\ee
Equation (\ref{eq:thetaTU}) follows from Eqn. 
(\ref{eq:antiproduct2}) of Appendix A for 
$\mathcal{O} = {\bf U_n}(\pi) {\bf T_a}$, 
$| \psi \rangle = | {\bf p} \rangle$, and 
$| \phi \rangle = {\bf U_n}(\pi) {\bf T_a} \theta | {\bf p} \rangle$ 
as soon as one observes, that 
$\left[ {\bf U_n}(\pi) {\bf T_a} \theta \right]^2 
 = {\bf T}_{\bf a}^2 = {\bf T}_{2 {\bf a}}$. 
In a magnetic field, double translation 
${\bf T}_{2 {\bf a}}$ remains a symmetry;  
according to the Bloch theorem, it acts 
on $| {\bf p} \rangle$ as per 
${\bf T}_{2 {\bf a}} | {\bf p} \rangle = 
e^{2i ({\bf p \cdot a})} | {\bf p} \rangle$, 
thus leading to (\ref{eq:thetaTU}).

Notice, however, that the eigenstate 
$ {\bf U_n}(\pi) {\bf T_a} \theta | {\bf p} \rangle$
carries momentum label $-{\bf p}$ rather than ${\bf p}$. 
By contrast with the case of zero field, combining 
${\bf U_n}(\pi)  {\bf T_a} \theta$ with inversion 
$\mathcal{I}$ no longer helps to produce a degenerate 
partner eigenstate at the original momentum ${\bf p}$: 
since $\theta$, ${\bf U_n} (\pi)$, and 
${\bf T_a} \mathcal{I}$ all commute, and since 
\mbox{$\left[ \mathcal{I} {\bf U_n}(\pi) {\bf T_a} 
\theta \right]^2=1$}, 
equation (\ref{eq:antiproduct2}) of Appendix A 
for $\mathcal{O} =  \mathcal{I} {\bf U_n}(\pi) {\bf T_a}$, 
$\psi = | {\bf p} \rangle$, and 
$\phi = \mathcal{I} {\bf U_n}(\pi) {\bf T_a} \theta 
 | {\bf p} \rangle$ only confirms, 
that $\langle {\bf p} |  \mathcal{I} 
{\bf U_n} (\pi) {\bf T_a} \theta | {\bf p}  \rangle$ 
equals itself. 

Thus, for an exact Bloch state $| {\bf p} \rangle$ 
at momentum ${\bf p}$, the anti-unitary symmetry 
${\bf U_n}(\pi) {\bf T_a} \theta$ produces an 
orthogonal degenerate eigenstate
${\bf U_n} (\pi) {\bf T_a} \theta
| {\bf p}  \rangle$ at momentum $-{\bf p}$.  
The two momenta ${\bf p}$ and $-{\bf p}$ 
are different, with one key exception. 
It occurs for ${\bf p}$ at the magnetic 
Brillouin zone boundary, given a unitary symmetry 
$\mathcal{U}$, that transforms $-{\bf p}$ into 
a momentum, equivalent to ${\bf p}$ up to a 
reciprocal lattice vector ${\bf Q}$ of the 
antiferromagnetic state \cite{dimmock2}: 
\be 
\label{eq:Dimmock_equivalence}
- \mathcal{U} {\bf p} = {\bf p} + {\bf Q}.
\ee
In this case, the eigenstate 
$\mathcal{U} {\bf U_n} (\pi)  {\bf T_a}
 \theta | {\bf p}  \rangle$ carries momentum 
label ${\bf p+Q \equiv p}$, is degenerate with 
$| {\bf p} \rangle$ and orthogonal to it, thus 
explicitly demonstrating Kramers degeneracy at 
momentum ${\bf p}$ in a transverse field. 
This result is general: combined with any 
momentum-inverting anti-unitary symmetry, equation 
(\ref{eq:Dimmock_equivalence}) leads to Kramers 
degeneracy at momentum ${\bf p}$ \cite{dimmock2}. 
The simplest illustration, where $\mathcal{U}$ 
is the unity o\-pe\-ra\-tor, is given by 
${\bf p} = {\bf Q}/2$ and shown in Figs. 
\ref{fig:BZ_1D} and \ref{fig:BZ_rectangular}(a) 
for two particular cases. These and other 
examples are described in Section IV. 
Notice that, at ${\bf p} = {\bf Q}/2$ 
(with $\mathcal{U} = {\bf 1}$), the degeneracy 
in a transverse field is guaranteed even for a 
low crystal symmetry, provided an inversion center. 

Also notice, once more, that $\mathcal{U}$, 
$\mathcal{I}$ and other point symmetries above 
% , that leave ${\bf \Delta_r}$ intact, 
are inert with respect to spin as a 
consequence of the `exchange symmetry' 
approximation \cite{andreev}. 

%%%%%%%%%% Figure %%%%%%%%
%%%%%%%%%%%%%%%%%%%%%%%%%%
\begin{figure}[h]
 \hspace{0cm}
 \epsfxsize=5cm
 \epsfbox{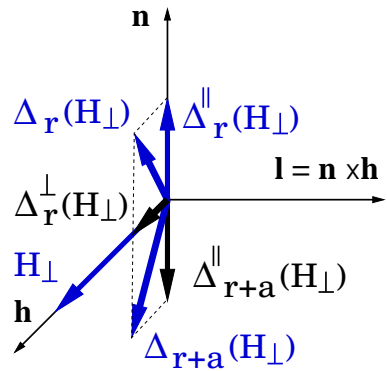}
 \vspace{15pt}
\caption{(color online). 
Relative orientation of ${\bf \Delta_r}(H_\perp)$ and 
${\bf \Delta_{r+a}}(H_\perp)$, tilted by a transverse 
field ${\bf H}_\perp$. 
The component $\bm{\Delta}_{\bf r}^\perp (H_\perp)$ 
along the field is periodic under elementary 
translation ${\bf T_a}$: 
$\bm{\Delta}_{\bf r+a}^\perp (H_\perp)
 = {\bf \Delta}_{\bf r}^\perp (H_\perp)$; 
the component along the zero-field  
${\bf \Delta_r}$ is anti-periodic: 
$\bm{\Delta}_{\bf r+a}^\| (H_\perp)
 = - \bm{\Delta}_{\bf r}^\| (H_\perp)$. 
} 
\label{fig:triad_canted}
\end{figure}
%%%%%%%%%%%%%%%%%%%%%%%%%%%
%%%%%%%%%%%%%%%%%%%%%%%%%%% 
Hamiltonian (\ref{eq:Hamiltonian_generic}) and the subsequent 
analysis ignored the response of the antiferromagnetic order 
to the transverse field ${\bf H}_\perp$. This, however, does 
not affect the set of points, where Kramers degeneracy in a 
transverse field is protected by the anti-unitary symmetry 
$\mathcal{U} {\bf U_n} (\pi)  {\bf T_a} \theta$. 
Upon application of ${\bf H}_\perp$, the N\'eel sublattices 
tilt towards the field, making it convenient to present 
${\bf \Delta_r}$ as 
\be
\label{eq:tilted_Delta}
{\bf \Delta_r}(H_\perp)
 = 
{\bf \Delta}_{\bf r}^\perp(H_\perp) + 
{\bf \Delta}_{\bf r}^\|(H_\perp), 
\ee
where $\bm{\Delta}_{\bf r}^\perp(H_\perp)$ 
points along ${\bf H}_\perp$, and 
$\bm{\Delta}_{\bf r}^\|(H_\perp)$ 
points along ${\bf n}$, as shown in Fig. 
\ref{fig:triad_canted}. 
Since $\bm{\Delta}_{\bf r+a}^\perp(H_\perp) 
=  \bm{\Delta}_{\bf r}^\perp(H_\perp)$ and 
$\bm{\Delta}_{\bf r+a}^\|(H_\perp) = 
 - \bm{\Delta}_{\bf r}^\|(H_\perp)$, the second 
column of Table \ref{table:symmetries} remains 
intact upon replacing ${\bf \Delta_r}$ 
in Hamiltonian (\ref{eq:Hamiltonian_generic}) 
by ${\bf \Delta_r}(H_\perp)$ 
of Eqn. (\ref{eq:tilted_Delta}).

If the antiferromagnetic unit cell is a multiple of its 
paramagnetic counterpart, the magnetic Brillouin zone 
boundary contains a set of points, that do not belong 
to the paramagnetic Brillouin zone boundary (for example, 
see Figs. \ref{fig:BZ_1D} and \ref{fig:BZ_rectangular}). 
In the paramagnetic state, no two points of this set,  
separated by antiferromagnetic reciprocal lattice vector 
${\bf Q}$ and satisfying condition 
(\ref{eq:Dimmock_equivalence}),
can be declared equivalent. As a curious consequence, 
the magnetic group of such a wave vector is {\em not} 
a subgroup of its paramagnetic counterpart \cite{dimmock2}. 
Hence the degeneracy, if present, 
does hinge on magnetic order. 

\section{III. Clues from weak coupling}

Additional insight into the locus of states, that 
remain degenerate in a transverse magnetic field, 
is afforded by a weak-coupling single-electron 
Hamiltonian in a doubly commensurate collinear  
antiferromagnet. Let ${\bf Q}$ be the 
antiferromagnetic ordering wave vector (see the 
examples below); ${\bf \Delta_r}$ creates a matrix 
element $({\bf \Delta} \cdot \bm{\sigma})$ between 
the Bloch states at momenta ${\bf p}$ and ${\bf p + Q}$ 
(for simplicity, I neglect its possible dependence 
on ${\bf p}$). Sublattice canting in a transverse 
field is taken into account in Appendix B. 
In magnetic field ${\bf H}$, and at weak coupling, 
Hamiltonian (\ref{eq:Hamiltonian_generic}) 
takes the form \cite{bralura} 
\be
\label{eq:weak_coupling_Hamiltonian}
\mathcal{H} 
 = 
\left[
\begin{array}{cc}
\epsilon_{\bf p}
 - ({\bf H} \cdot \bm{\sigma}) 
 & 
({\bf \Delta} \cdot \bm{\sigma}) \\
 & \\
({\bf \Delta} \cdot \bm{\sigma}) 
 & 
\epsilon_{\bf p + Q}
 - ({\bf H} \cdot \bm{\sigma}) 
\end{array}
\right]
,
\ee
where 
$\epsilon_{\bf p}$  and $\epsilon_{\bf p + Q}$ 
are single-particle energies of $\mathcal{H}_0$ 
in (\ref{eq:Hamiltonian_generic}) at momenta 
${\bf p}$ and ${\bf p + Q}$, and the `bare' 
$g$-tensor in $({\bf H} \cdot \bm{\sigma})$ 
is omitted for brevity. 

In a purely transverse field ${\bf H}_\perp$, 
this Hamiltonian can be diagonalized simply by 
choosing the $\hat{z}$-axis in spin space along 
${\bf H}_\perp$, and the $\hat{x}$-axis along 
$\bm{\Delta}$. As a result, Hamiltonian 
(\ref{eq:weak_coupling_Hamiltonian}) splits 
into two decoupled pieces: 
$\mathcal{H}_1 ({\bf p}, {\bf H}_\perp)$ 
for the amplitudes $| \bf{p} ; \uparrow \rangle$ 
and $| \bf{p + Q} ; \downarrow \rangle$, and 
$\mathcal{H}_2 ({\bf p}, {\bf H}_\perp)$ 
for the amplitudes 
$| \bf{p} ; \downarrow \rangle$ and 
$| \bf{p + Q} ; \uparrow \rangle$: 
\be
\label{eq:weak_coupling_Hamiltonian_decoupled}
\mathcal{H}_{1(2)} 
({\bf p}, {\bf H}_\perp) 
 = 
\left[
\begin{array}{cc}
\epsilon_{\bf p} \mp H_\perp & \Delta \\
 & \\
 \Delta
 & 
\epsilon_{\bf p + Q} \pm H_\perp 
\end{array}
\right].
\ee
The spectra $\mathcal{E}_{1(2)}({\bf p})$ 
of $\mathcal{H}_{1(2)}$ are given by  
\be
\label{eq:weak_coupling_spectrum}
\mathcal{E}_{1(2)}({\bf p}, {\bf H}_\perp)
 = 
\eta_{\bf p} 
            \pm
\sqrt{
\Delta^2
 +
\left[
\zeta_{\bf p} \mp  
({\bf H}_\perp \cdot \bm{\sigma}) 
\right]^2
},
\ee  
with $({\bf H}_\perp \cdot \bm{\sigma}) = H_\perp$ 
corresponding to $\mathcal{H}_1$, and 
$({\bf H}_\perp \cdot \bm{\sigma}) = - H_\perp$ to 
$\mathcal{H}_2$, and with 
$\eta_{\bf p} \equiv 
\frac{\epsilon_{\bf p} + \epsilon_{\bf p + Q}}{2}
$, 
and 
$
\zeta_{\bf p} \equiv 
\frac{\epsilon_{\bf p} - \epsilon_{\bf p + Q}}{2}
$.
The same spectrum can be obtained by excluding, 
say, $| \bf{p+Q} ; \sigma \rangle$ from the eigenvalue 
equation for (\ref{eq:weak_coupling_Hamiltonian}), 
but it is important to keep in mind that 
$\bm{\sigma}$ in (\ref{eq:weak_coupling_spectrum}) 
no longer describes spin, but rather pseudospin: 
since $({\bf H}_\perp \cdot \bm{\sigma})$ does 
not commute with the Hamiltonian, the eigenstates 
of $\mathcal{H}_{1(2)}$ are superpositions of 
spin-up and spin-down states. 

Equation (\ref{eq:weak_coupling_spectrum}) illustrates 
a number of points. Firstly, the electron spectrum 
acquires a gap of size $2\Delta$. Se\-cond\-ly, in the 
absence of magnetic field, each eigenstate is indeed 
doubly degenerate, in agreement with the arguments, 
encapsulated in Eqn. (\ref{eq:Kramers}). 
Thirdly, Eqn. (\ref{eq:weak_coupling_spectrum}) shows, 
that the degeneracy persists in a transverse field 
(and, therefore, $g_\perp({\bf p})$ 
in Eqn. (\ref{eq:ZSO}) vanishes) 
whenever $\zeta_{\bf p} = 0$. 
Barring a special situation, this equation 
defines a surface in three dimensions, a line 
in two, and a set of points in one. This result 
for the dimensionality of the manifold of 
degenerate states hinges solely on the symmetry 
of the antiferromagnetic state and holds beyond 
weak coupling, as shown in Appendix C. 
Furthermore, as shown above, this ma\-ni\-fold 
must contain all points, satisfying 
Eqn. (\ref{eq:Dimmock_equivalence}): 
the points, where the degeneracy is enforced 
by symmetry. Finally, expansion of Eqn. 
(\ref{eq:weak_coupling_spectrum}) to first order in 
$(\bf{H}_\perp \cdot \bm{\sigma})$ yields the expression 
for $g_\perp({\bf p})$ in (\ref{eq:ZSO}) within the weak 
coupling model (\ref{eq:weak_coupling_Hamiltonian}): 
\be
\label{eq:weak_coupling_g} 
g_\perp({\bf p}) = 
\frac{\zeta_{\bf p}}{\sqrt{\Delta^2 + \zeta_{\bf p}^2}}.
\ee

At the end of the preceding Subsection, I showed that 
tilting of the N\'eel sublattices in a transverse field 
does not affect the set of points, where Kramers degeneracy 
in a transverse field is protected by the anti-unitary 
symmetry $\mathcal{U} {\bf U_n} (\pi)  {\bf T_a} \theta$. 
However, generally, the rest of the degeneracy manifold 
is not protected by symmetry, and may change shape upon 
crystal deformation, or under a\-no\-ther perturbation. 
For instance, while leaving intact the 
symmetry-protected set of degeneracy points, 
the sublattice canting may change the shape 
of the degeneracy manifold $g_\perp({\bf p}) = 0$ 
compared with $\zeta_{\bf p} = 0$. 
This effect is discussed in Appendix B. 

Put otherwise, the degeneracy manifold may be 
divided into two parts. The first part is the 
`Kramers degeneracy' subset of special momenta, 
fixed by conspiracy between the anti-unitary symmetry 
${\bf U_n} (\pi)  {\bf T_a} \theta$ and the crystal 
symmetry. 
This `Kramers' subset is insensitive to perturbations  
that leave intact the crystal symmetry of the material. 
The rest is an `accidental' degeneracy subset, whose 
geometry, by contrast, may vary under perturbations, 
that do not affect the crystal symmetry, but only 
alter the microscopic parameters of the system. 
The division of the degeneracy manifold into the 
`Kramers' and the `accidental' degeneracy subsets 
is well illustrated by the examples of two-dimensional 
rectangular and square symmetry antiferromagnets 
in Section IV.

\subsection{Spectral symmetries in momentum space}

The spectrum of Hamiltonian  
(\ref{eq:weak_coupling_Hamiltonian}) enjoys a 
number of symmetries. Firstly, inversion symmetry 
makes the spectrum even under inversion. At the same time, 
$g_\perp ({\bf p}) ({\bf H}_\perp \cdot \bm{\sigma})$ 
must also be even under inversion, which implies 
\be
\label{eq:g-even}
g_\perp (-{\bf p}) = g_\perp ({\bf p}).
\ee
The anti-unitary symmetry ${\bf U_l}(\pi) \theta$ 
in the last line of Table \ref{table:symmetries} 
is another reason for $g_\perp ({\bf p})$ to be 
even under inversion, as ${\bf U_l}(\pi) \theta$ 
turns $g_\perp ({\bf p}) ({\bf H}_\perp \cdot \bm{\sigma})$ 
into $g_\perp (-{\bf p}) ({\bf H}_\perp \cdot \bm{\sigma})$.

\begin{figure}[h]
\centerline{
\mbox{\includegraphics[width=1.6in]{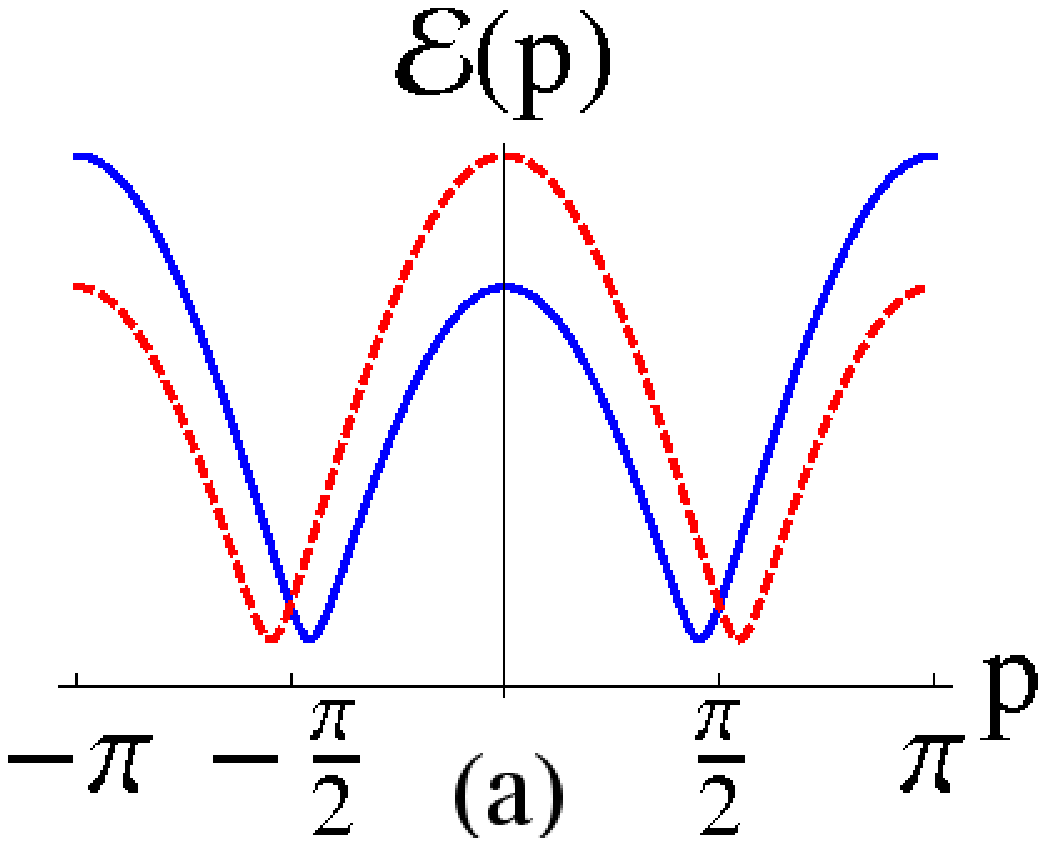}}
 \hspace{-0.cm}
\mbox{\includegraphics[width=1.6in]{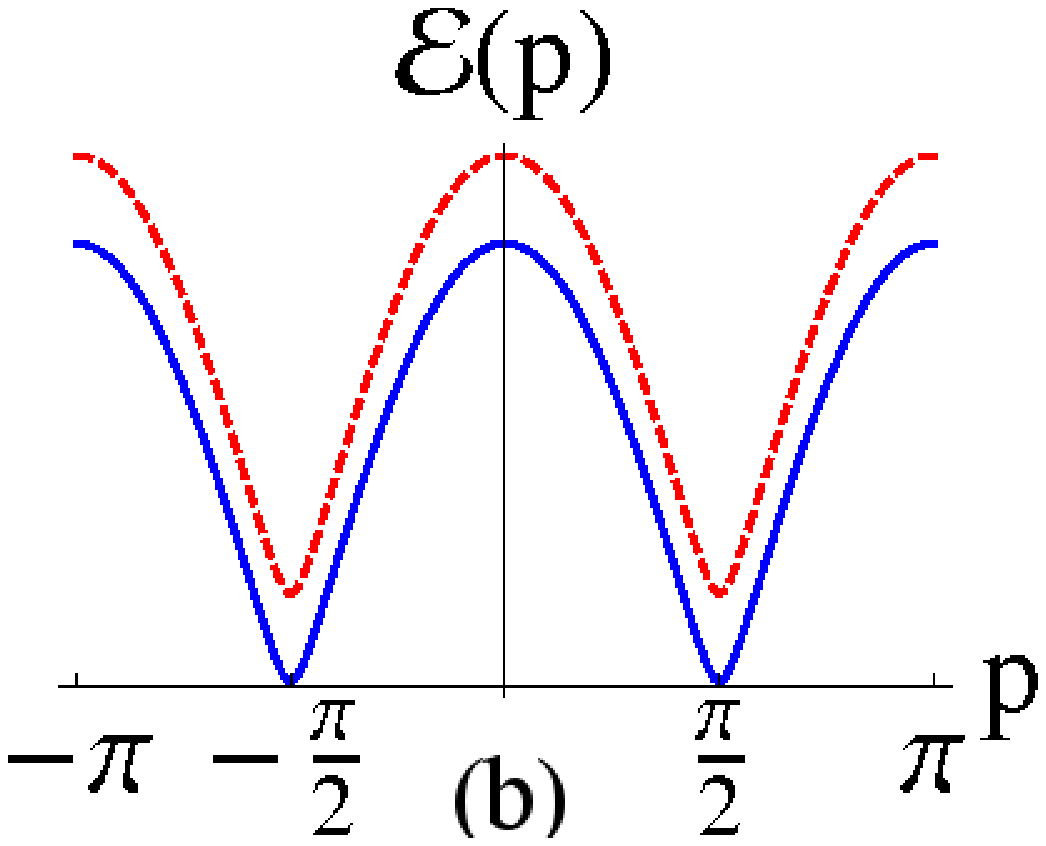}}
 \hspace{0.2cm}
 }
 \caption{(color online). 
One-dimensional conduction band splitting. 
(a) Conduction band [`+' sign in front of 
the square root in Eqn. 
(\ref{eq:weak_coupling_spectrum})], 
split by a transverse field. Here Eqn. 
(\ref{eq:Q-boost_spectrum}) is illustrated 
by spectrum (\ref{eq:weak_coupling_spectrum}) 
for a nearest-neighbor hopping 
$\epsilon_{\bf p} = t \cos p$ in Hamiltonian 
(\ref{eq:weak_coupling_Hamiltonian}). Notice 
that, in spite of period doubling in real space, 
neither of the two split spectra has period $\pi$; 
instead, both are $2\pi$-periodic, but map 
onto each other upon translation by $\pi$. 
Also notice that the two split sub-bands intersect 
at the symmetry-enforced degeneracy point 
${\bf p}=\pm \frac{\pi}{2}$, as they should 
(see the one-dimensional example of the 
following section). 
(b) Conduction band, split by a longitudinal 
field. By contrast with (a), each split sub-band 
is $\pi$-periodic, and degeneracy is lifted for 
all momenta.
}
\label{fig:transverse_splitting}
\end{figure}

Periodicity doubling due to antiferromagnetism 
ma\-ni\-fests itself more interestingly. The 
ordering wave vector ${\bf Q}$ being a reciprocal 
lattice vector, any Bloch eigenstate at momentum 
${\bf p}$ must have a degenerate partner eigenstate 
at momentum ${\bf p+Q}$. Usually, this implies 
${\bf Q}$-periodicity of any given band: 
\mbox{$\epsilon({\bf p}) = \epsilon({\bf p + Q})$}, 
which is the case, for instance, in a longitudinal 
field ${\bf H}_\|$, where 
$\mathcal{E}({\bf p})$ undergoes the common Zeeman 
splitting 
\mbox{
$\mathcal{E}({\bf p}) \rightarrow 
\mathcal{E}({\bf p}) \pm H_\|$}. 
In a N\'eel antiferromagnet in a transverse field, 
this is not the case: for a general ${\bf p}$, 
\mbox{$\mathcal{E}_{1(2)}({\bf p + Q})
 \neq \mathcal{E}_{1(2)}({\bf p})$}. 
Instead, 
\be
\label{eq:Q-boost_spectrum} 
\mathcal{E}_1({\bf p + Q}, {\bf H}_\perp)
 = 
\mathcal{E}_2({\bf p}, {\bf H}_\perp), 
\ee
while both $\mathcal{E}_1({\bf p})$ and 
$\mathcal{E}_2({\bf p})$ 
of Eqn. (\ref{eq:weak_coupling_spectrum}) 
are invariant under 
momentum shift ${\bf p} \rightarrow {\bf p}+2{\bf Q}$.  
These properties are illustrated in 
Fig. \ref{fig:transverse_splitting}, 
showing the splitting of a one-dimensional 
conduction band in a transverse field. 

%%%%%%% Figure %%%%%%%
%%%%%%%%%%%%%%%%%%%%%%
The reason behind (\ref{eq:Q-boost_spectrum}) 
is that, as long as $H_\perp \neq 0$, neither 
$\mathcal{H}_1$ nor $\mathcal{H}_2$ in Eqn.  
(\ref{eq:weak_coupling_Hamiltonian_decoupled}) 
is invariant under the momentum boost  
$\bf{p \rightarrow p+Q}$, in spite of the 
Hamiltonians (\ref{eq:Hamiltonian_generic}) 
and (\ref{eq:weak_coupling_Hamiltonian}) 
both having doubled periodicity. Rather, 
\be
\label{eq:Q-boost_Hamiltonian} 
\mathcal{H}_1({\bf p + Q}, {\bf H}_\perp)
 = 
\mathcal{H}_2({\bf p}, {\bf H}_\perp), 
\ee
which is made explicit by subsequent 
exchange of the diagonal matrix elements, 
and leads to Eqn. (\ref{eq:Q-boost_spectrum}). 

Zeeman splitting corresponds to the difference 
\mbox{$\mathcal{E}_1({\bf p}, {\bf H}_\perp)
 - \mathcal{E}_2({\bf p}, {\bf H}_\perp)$}; 
hence it changes sign upon momentum shift by $\bf{Q}$.
At the same time, direct inspection shows that 
Hamiltonians $\mathcal{H}_1$ and $\mathcal{H}_2$ in  
Eqn. (\ref{eq:weak_coupling_Hamiltonian_decoupled}) 
turn into one another upon inversion of $\bf{H}_\perp$: 
\mbox{$\mathcal{H}_1 ({\bf p}, -{\bf H}_\perp) = 
 \mathcal{H}_2 ({\bf p}, {\bf H}_\perp)$}. Combining 
this with Eqn. (\ref{eq:Q-boost_Hamiltonian}), one 
finds that momentum boost by $\bf{Q}$ accompanied 
by inversion of $\bf{H}_\perp$ is a symmetry 
of both $\mathcal{H}_1$ and $\mathcal{H}_2$: 
\be 
\label{eq:my_symmetry}
\mathcal{H}_{1(2)} ({\bf p+Q}, -{\bf H}_\perp) 
 = 
\mathcal{H}_{1(2)} ({\bf p}, {\bf H}_\perp).
\ee
For the transverse Zeeman term 
$g_\perp ({\bf p}) ({\bf H}_\perp \cdot \bm{\sigma})$,  
this yields  
\be
\label{eq:Q-boost_for_g}
g_\perp ({\bf p + Q}) = - g_\perp ({\bf p}).
\ee
Combined, Eqns. (\ref{eq:g-even}) 
and (\ref{eq:Q-boost_for_g}) lead to 
\be 
\label{eq:around_Q/2}
g_\perp (\frac{\bf Q}{2} + {\bf p})
 = - g_\perp (\frac{\bf Q}{2} - {\bf p}).
\ee
This implies not only that $g_\perp ({\bf p})$ must 
vanish at ${\bf p} = \frac{\bf Q}{2}$, but also 
that $g_\perp (\frac{\bf Q}{2} + {\bf p})$ is an 
odd function of ${\bf p}$. The conclusions of this 
subsection hold after the sublattice tilting 
is taken into account (see Appendix B). 

\section{IV. Examples} 

In this section, I describe the manifolds of 
degenerate states for a number of concrete 
examples and thus show, that the Zeeman 
spin-orbit coupling (\ref{eq:ZSO}) is at work 
in many materials of great interest. It gives 
rise to va\-ri\-ous interesting phenomena, 
some of which are outlined in the subsection 
`Experimental signatures' of Section V. 

Kramers degeneracy in a transverse field and the 
Zeeman spin-orbit coupling (\ref{eq:ZSO}) will  
manifest themselves whenever carriers are 
present at or near the manifold of degenerate 
states $g_\perp ({\bf p}) = 0$. In a weakly 
doped antiferromagnetic insulator, this will 
happen whenever the relevant band extremum 
falls at or near the manifold of degenerate 
states. In an antiferromagnetic metal, this 
occurs when the Fermi surface crosses this 
manifold. Hence, for metals, I mention the 
Fermi surface geometry whenever known. 

Between these two limiting cases of a weakly 
doped antiferromagnetic insulator and an 
antiferromagnetic metal with a large Fermi 
surface, the experimental manifestations of 
the Zeeman spin-orbit coupling will be 
quantitatively different. On top of this, 
certain effects will be sensitive to the 
geometry of the degeneracy manifold and 
its intersection with the Fermi surface, 
as well as the orientation of the staggered 
magnetization with respect to the crystal 
axes. A detailed discussion of these effects 
will be presented elsewhere. 

When selecting the examples below, the preference 
was given to materials, available in high-purity 
samples, where de Haas-van Alphen oscillations 
were observed, and where magnetic structure was 
unambiguously cha\-rac\-te\-rized by neutron 
scattering. As explained in the Introduction, 
the results of this work apply to materials well 
inside a long-range antiferromagnetic phase, and 
far enough from any critical point, quantum or 
classical. For both quantum and thermal 
fluctuations of antiferromagnetic order to 
be negligible, the ordered moment shall be 
noticeable on the scale of the Bohr magneton, 
and the sample shall be kept well below 
both the N\'eel and the effective Fermi 
temperatures.

\subsection{One dimension}
\begin{figure}[h]
 \hspace{-0.5cm}
 \epsfxsize=4cm
 \epsfbox{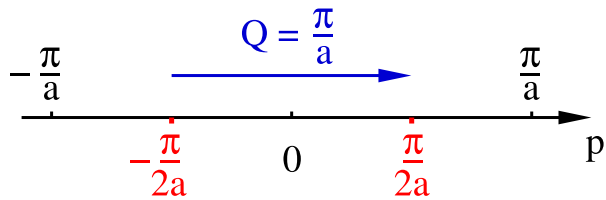}
 \vspace{15pt}
\caption{(color online). 
The paramagnetic (${\bf p} = \pm\frac{\pi}{a}$), and 
the antiferromagnetic (${\bf p} = \pm\frac{\pi}{2a}$) 
Brillouin zone boundaries of a one-dimensional N\'eel 
antiferromagnet. In the antiferromagnetic state, 
the two points ${\bf p} = \pm\frac{\pi}{2a}$ 
are identical up to the reciprocal lattice vector 
${\bf Q} = \frac{\pi}{a}$ of the antiferromagnetic 
state. At these two points, anti-unitary symmetry 
${\bf U_n}(\pi){\bf T_a} \theta$ protects the Kramers 
degeneracy against transverse magnetic field. 
} 
\label{fig:BZ_1D}
\end{figure}

In one dimension, the magnetic Brillouin zone 
boun\-da\-ry reduces to two points 
${\bf p} = \pm \frac{\pi}{2a}$, which in fact 
coincide up to the antiferromagnetic wave 
vector ${\bf Q} = \frac{\pi}{a}$, that is also 
the reciprocal lattice vector of the antiferromagnetic 
state (see Fig. \ref{fig:BZ_1D}). In terms of the 
general condition (\ref{eq:Dimmock_equivalence}), 
this is the simplest case: $\mathcal{U} = {\bf 1}$. 

As a result, at ${\bf p} = \pm \frac{\pi}{2a}$, 
the two exact Bloch states in a transverse field, 
$| {\bf p} \rangle$ and 
$\theta {\bf T_a U_n} (\pi) | {\bf p} \rangle$, 
correspond to the {\em same} momentum ${\bf p}$, and 
are degenerate by virtue of $\theta {\bf T_a U_n} (\pi)$ 
being a symmetry. Equation (\ref{eq:thetaTU}) 
guarantees their orthogonality, thus protecting 
Kramers degeneracy at momentum ${\bf p} = \pm \frac{\pi}{2a}$ 
against transverse magnetic field. 
%%%%%%%%%%%%%%%%%%%%%
%
% \includegraphics[width=60mm]{fig1.pdf}
% 
%%%%%%%%%%%%%%%%%%%%% Figure %%%%%%%%%%%%%%%%%%%%%%%%%%
%%%%%%%%%%%%%%%%%%%%%%%%%%%%%%%%%%%%%%%%%%%%%%%%%%%%%%

\subsection{Two dimensions, rectangular and square symmetries}

\begin{figure}[h]
 \hspace{3cm}
 \epsfxsize=8cm
 \epsfbox{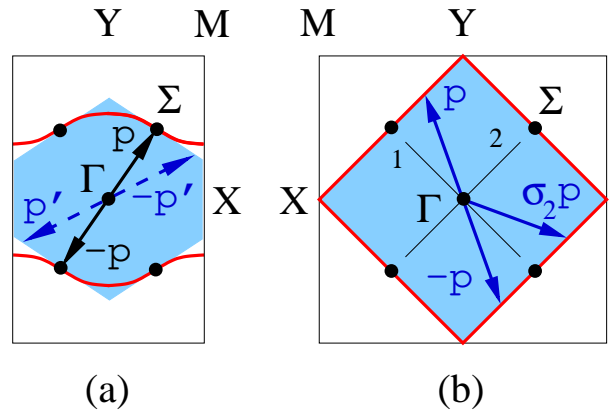}
 \vspace{15pt}
\caption{(color online). 
Geometry of the problem. 
(a) The Brillouin zone for a simple rectangular 
lattice (the rectangle), and its antiferromagnetic 
counterpart (MBZ, shaded hexagon). Thick (red) curve, 
passing through point $\Sigma$, shows a typical 
degeneracy line $g_\perp ({\bf p}) = 0$. 
At the MBZ boundary, only momentum ${\bf p}$ 
at point $\Sigma$ is equivalent to ${- \bf p}$ 
up to a reciprocal lattice vector of the 
antiferromagnetic state; for a generic ${\bf p'}$, 
shown by the dashed arrow, this is not true. 
(b) The Brillouin zone of a simple square lattice 
and  its antiferromagnetic counterpart (shaded 
diagonal square). The degeneracy line must contain 
the entire MBZ boundary, shown in red. Point $\Gamma$ 
is the Brillouin zone center, points $X$ and $Y$ lie 
at the centers of the paramagnetic Brillouin Zone edges. 
Point $\Sigma$ lies at the center of the MBZ boundary. 
} 
\label{fig:BZ_rectangular}
\end{figure}

Now consider a two-dimensional antiferromagnet 
on a lattice of rectangular or square symmetry, 
with the or\-de\-ring wave vector ${\bf Q}=(\pi,\pi)$. 
In a transverse magnetic field, the degeneracy 
persists on a line in the Brillouin zone, by virtue 
of Eqn. (\ref{eq:weak_coupling_spectrum}). 
I will show that, in the rectangular case, the 
degeneracy line must contain the point $\Sigma$ 
at the center of the magnetic Brillouin zone (MBZ) 
boundary (i.e. the star of ${\bf p} = {\bf Q}/2$), 
shown in Fig. \ref{fig:BZ_rectangular}(a). 
In the square symmetry case, the degeneracy persists at 
the entire MBZ boundary (Fig. \ref{fig:BZ_rectangular}(b)). 
The MBZ in Fig. \ref{fig:BZ_rectangular} is the reciprocal  
space counterpart of the Wigner-Seitz cell of the magnetic state (Fig. \ref{fig:AF_real_space}), and the ordering 
wave vector ${\bf Q}=(\pi,\pi)$ connects points $X$ and 
$Y$ in Figs. \ref{fig:BZ_rectangular}(a) and (b). 
%%%%%%%%%%%%%%%%%%%%%
%
% \includegraphics[width=60mm]{fig1.pdf}
% 
%%%%%%%%%%%%%%%%%%%%% Figure %%%%%%%%%%%%%%%%%%%%%%%%%%

%%%%%%%%%%%%%%%%%%%%%%%%%%%%%%%%%%%%%%%%%%%%%%%%%%%%%%
Consider a Bloch state $| {\bf p} \rangle$ 
at momentum ${\bf p}$ in a transverse field. 
As discussed in Section II, the eigenstate 
$ \theta {\bf T_a U_n}(\pi) | {\bf p} \rangle$ at 
momentum $-{\bf p}$ is degenerate with $| {\bf p} \rangle$ 
and, according to Eqn. (\ref{eq:thetaTU}), must be 
orthogonal to it unless $({\bf p} \cdot {\bf a})$ 
is an integer multiple of $\pi$ -- put otherwise, 
unless ${\bf p}$ belongs to the paramagnetic 
Brillouin zone boundary. At points $\Sigma$, X, 
and Y, momenta ${\bf p}$ and $-{\bf p}$ coincide up 
to a reciprocal lattice vector of the antiferromagnetic 
state. However, at points X and Y (as well as at the 
entire vertical segment of the MBZ boundary 
in Fig.  \ref{fig:BZ_rectangular}(a)), 
$({\bf p} \cdot {\bf a})$ is an integer multiple 
of $\pi$; hence $| {\bf p} \rangle$ and 
$\theta {\bf T_a U_n}(\pi) | {\bf p} \rangle$ 
are not obliged to be orthogonal  there as per 
Eqn. (\ref{eq:thetaTU}). Thus, $\Sigma$ 
is the only point at the MBZ boundary, where the 
two degenerate states $| {\bf p} \rangle$ and 
$ \theta {\bf T_a U_n}(\pi) | {\bf p} \rangle$ 
are orthogonal and correspond to the \textit{same} 
momentum: Dashed arrows in figure 
\ref{fig:BZ_rectangular}(a) show, that, for a 
generic point ${\bf p}'$ at the MBZ boundary, 
no symmetry operation relates $-{\bf p}'$ to a vector, 
equivalent to ${\bf p}'$. Therefore, it is only at 
point $\Sigma$, that the symmetry protects Kramers 
degeneracy against transverse magnetic field. 
As in the one-dimensional example of the previous 
subsection, in terms of Eqn. 
(\ref{eq:Dimmock_equivalence}) this corresponds 
to the simplest case of $\mathcal{U}={\bf 1}$.

This can be illustrated by a nearest-neighbor 
hopping spectrum 
\be 
\label{eq:hopping}
\epsilon_{\bf p} = t 
\left[ 
\cos p_x + \eta \cos p_y
\right]
\ee 
in the weak-coupling example of the previous subsection: 
for rectangular symmetry ($\eta \neq 1$), spectrum 
(\ref{eq:weak_coupling_spectrum}) in a transverse field 
remains degenerate at a thick (red) line, sketched in 
Fig. \ref{fig:BZ_rectangular}(a). Upon variation 
of $\eta \neq 1$, the line changes its shape, 
but remains pinned at the star of wave vector 
${\bf p = Q}/2$ (i.e. at point $\Sigma$) 
in Fig. \ref{fig:BZ_rectangular}(a). In terms of the 
preceding subsection, the star of ${\bf p = Q}/2$ is 
the `Kramers' subset of the degeneracy manifold,  
while the rest of the degeneracy line in Fig. 
\ref{fig:BZ_rectangular}(a) is the `accidental' 
degeneracy subset. 

Promotion from the rectangular symmetry 
\mbox{($\eta \neq 1$)} to that of a square 
\mbox{($\eta = 1$)} brings along invariance 
under reflections $\sigma_{1, 2}$ in 
% (or three-dimensional coordinate
% rotations by $\pi$ around) 
either of the two diagonal axes $1$ and $2$, 
passing through point $\Gamma$ 
in Fig. \ref{fig:BZ_rectangular}(b). 
As a result, the eigenstate 
$\sigma_1 \theta {\bf T_a U_n}(\pi) | {\bf p} \rangle$ 
at momentum $\sigma_2 {\bf p}$ 
(Fig. \ref{fig:BZ_rectangular}(b)) is also degenerate 
with $| {\bf p} \rangle$ and orthogonal to it, as one 
can show analogously to the examples above. In terms 
of the ge\-ne\-ral condition (\ref{eq:Dimmock_equivalence}), 
this means $\mathcal{U} = \sigma_{1, 2}$.

A momentum ${\bf p}$ at the MBZ boundary in 
Fig. \ref{fig:BZ_rectangular}(b) differs from 
$\sigma_2 {\bf p}$ by a reciprocal lattice vector; 
thus the two momenta coincide in the nomenclature 
of the antiferromagnetic Brillouin zone. Hence, for 
a square-symmetry lattice in a transverse field, 
the degeneracy is of a `Kramers' (i.e. symmetry-protected) 
nature at the entire MBZ boundary, as shown in Fig.  
\ref{fig:BZ_rectangular}(b). In this case, barring 
a particularly pathological band structure, 
the degeneracy manifold is exhausted by its 
`Kramers' subset. 

In accordance with the symmetry arguments above, 
for the toy nearest-neighbor hopping spectrum  
(\ref{eq:hopping}) at the square symmetry point 
$\eta=1$, the degeneracy line of Eqn. (\ref{eq:weak_coupling_spectrum}) coincides 
with the MBZ boundary, as shown in Fig. 
\ref{fig:BZ_rectangular}(b). 
By contrast, for rectangular symmetry, it is Eqn. (\ref{eq:weak_coupling_spectrum}) that restricts  
the degeneracy in a transverse field to a line in momentum 
space, and it is the symmetry that pins this line at point 
$\Sigma$ at the middle of the MBZ boundary, as shown in 
Fig. \ref{fig:BZ_rectangular}(a).

Now, $g_\perp ({\bf p})$ can be expanded 
in a vicinity of the degeneracy line  
$g_\perp ({\bf p}) = 0$. With the exception 
of higher-symmetry points, such as point $X$ 
in Fig. \ref{fig:BZ_rectangular}(b), the leading 
term of the expansion is linear in momentum 
deviation ${\bf \delta p}$ from the degeneracy 
line: 
\be
g_\perp ({\bf p})
   \approx 
\frac{{\bf \Xi}_{\bf p} \cdot \delta {\bf p}}{\hbar}, 
\label{eq:expand_g}
\ee 
where ${\bf \Xi}_{\bf p}/\hbar$ is the momentum 
gradient of $g_\perp ({\bf p})$ at point ${\bf p}$ 
on the degeneracy line. As mentioned in the previous 
Section, inversion symmetry makes $g_\perp ({\bf p})$ 
even under inversion. Therefore, ${\bf \Xi_{\bf p}}$ 
changes sign upon inversion, which is consistent with Eqns. 
(\ref{eq:Q-boost_spectrum}) and (\ref{eq:Q-boost_for_g}), 
that require $g_\perp({\bf p})$ to change sign upon 
momentum shift by ${\bf Q}$. 

As shown in the last subsection of Section III,  
$g_\perp ({\bf p})$ is an odd function of the 
deviation $\delta {\bf p}$ from point 
$\Sigma$ (the star of ${\bf p} = {\bf Q}/2$) 
in Figs. \ref{fig:BZ_rectangular}(a) and (b). 
Therefore, expansion of $g_\perp ({\bf p})$ 
around point $\Sigma$ cannot contain an even 
power of $\delta {\bf p}$.  

\subsection{Chromium}

This subsection is devoted to commensurate 
antiferromagnetism in chromium -- the simplest 
of magnetic orders, occurring in this textbook 
spin density wave metal. Chromium crystallizes 
into a b.c.c. lattice, and undergoes various  
magnetic and structural transitions upon 
variation of temperature, pressure, or alloying \cite{fawcett_1,fawcett_2,kulikov}. 
%%%%%%%%%% Figure %%%%%%%%%% %%%%%%%%%%%%%%%%%%%%%%%%
\begin{figure}[h]
 \hspace{0cm}
 \epsfxsize=3cm
 \epsfbox{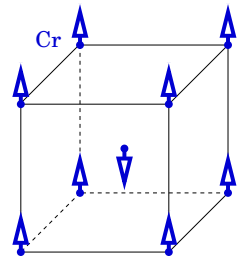}
 \vspace{15pt}
\caption{(color online). 
Schematic drawing of the commensurate magnetic 
structure of antiferromagnetic chromium. 
} 
\label{fig:Cr_real_space}
\end{figure}
%%%%%%%%%%%%%%%%%%%%%%
%%%%%%%%%%%%%%%%%%%%%%
%%%%%%% Figure %%%%%%%
\begin{figure}[h]
\centerline{
   \mbox{\includegraphics[width=1.8in]{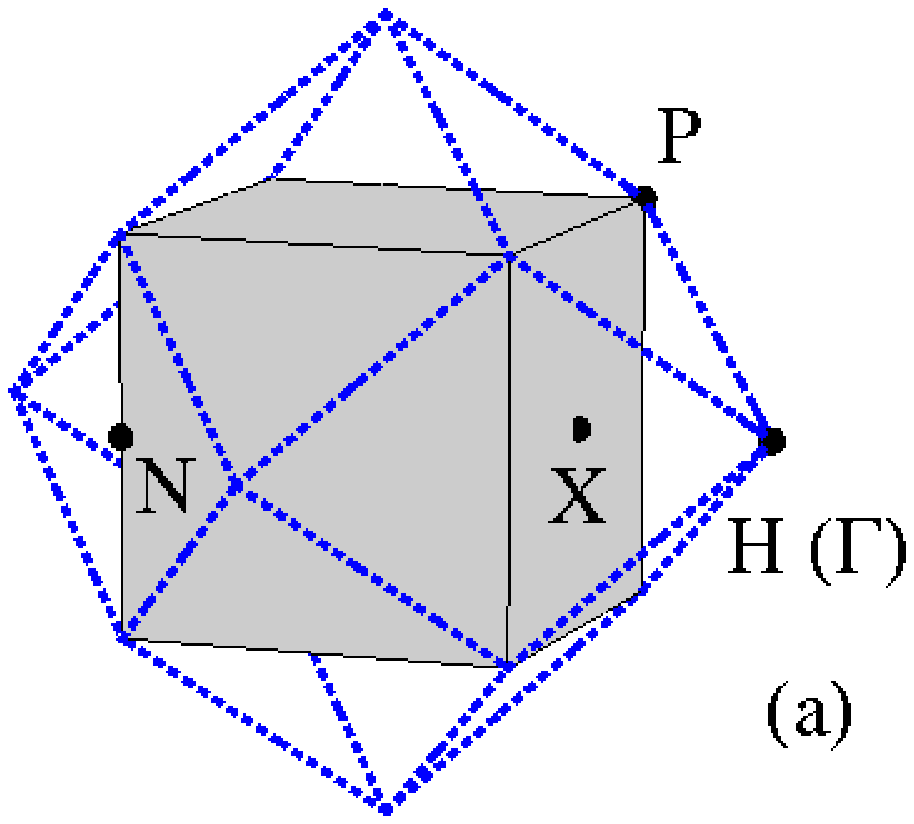}}
 \hspace{-0.cm}
   \mbox{\includegraphics[width=1.6in]{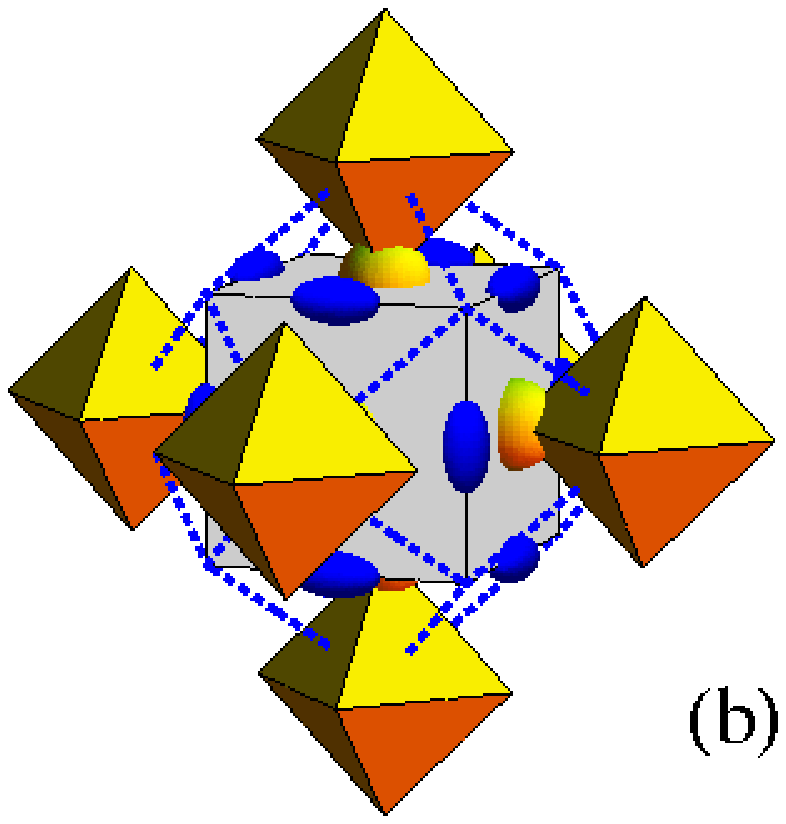}}
 \hspace{0.2cm}
 }
 \caption{(color online). 
The Brillouin zone and the Fermi surface of 
commensurate antiferromagnetic chromium. 
(a) Dashed lines show the paramagnetic 
Brillouin zone boun\-da\-ry. The solid cube 
inside it is the magnetic Brillouin zone (MBZ), 
with high symmetry points indicated. In the 
antiferromagnetic state, point $H$ is equivalent 
to point $\Gamma$ at the center of the Brillouin 
zone (not shown). 
(b) The Fermi surface sketch of paramagnetic chromium. 
As in (a), dashed lines show the paramagnetic Brillouin 
zone boundary, and the solid cube inside it is the MBZ. 
The nearly octahedral hole Fermi surface is centered 
at point $H$, and nearly spherical electron `balls' 
are located at face centers $X$ of the MBZ boundary. 
Together with the nearly octahedral electron surface, 
centered at point $\Gamma$ (not shown), these electron 
balls form the electron `jack'. A set of hole ellipsoids 
is centered at points $N$ in the middle of the magnetic Brillouin zone edges.
}
 \label{fig:Cr}
\end{figure}
%%%%%%%%%%%%%%%%%%%%%%

Below the N\'eel temperature $T_N$ of about 311 K 
at ambient pressure, chromium develops weakly 
incommensurate antiferromagnetism with ordered 
moment of about 0.5 $\mu_B$ per atom at 4.2K. 
However, strain -- or doping with some 0.1 to 
0.3$\%$ of a transition metal (such as Mn, Re, 
Rh, Ru, Ir, Os or Pt \cite{fawcett_2}) 
 -- eliminate incommensurability in favor of 
commensurate order with wave vector $\left[ 0 0 1 \right]$, 
shown in Fig. \ref{fig:Cr_real_space}. Commensurate 
order has also been observed and much studied in 
thin films of chromium, often with an enhanced 
N\'eel temperature and ordered moment \cite{zabel}. 
This article neglects fluctuations of magnetic 
order and, conveniently, the high N\'eel 
temperature of chromium facilitates experimental 
access to $T \ll T_N$, where thermal fluctuations 
are suppressed. 

The paramagnetic and the antiferromagnetic Brillouin 
zones for bulk commensurate antiferromagnetic chromium 
are shown in Fig. \ref{fig:Cr}(a). 
An arbitrary momentum at the MBZ boundary becomes 
equivalent to its opposite upon reflection in a  
properly chosen plane. 
Si\-mi\-lar\-ly to the two-dimensional square-symmetry 
e\-xam\-ple above, this equivalence is up to a primitive 
wave vector of the antiferromagnetic reciprocal space. 
Hence, in a transverse magnetic field, the Kramers 
degeneracy survives at the entire magnetic Brillouin 
zone boundary in Fig. \ref{fig:Cr}(a). 

The disappearance of $g_\perp ({\bf p})$ affects 
electrons at two different sheets of the Fermi 
surface, sketched in Fig. \ref{fig:Cr}(b): 
those at the nearly spherical electron parts, centered 
at points $X$ in the middle of each MBZ face, and those 
at the hole ellipsoids, centered at points $N$ in the 
middle of each MBZ edge. 
For the former, the leading term of the expansion is linear 
in the momentum deviation $\delta p_\perp$ from the flat 
face of the MBZ boundary. For the latter, the leading 
term of the expansion is quadratic near each MBZ edge, 
since $g_\perp ({\bf p})$ vanishes at each of the two 
intersecting faces of the MBZ boundary. 

\subsection{CeIn$_3$, UIn$_3$, UGa$_3$ ...} 

A number of cerium and uranium binary intermetallics of 
simple cubic Cu$_3$Au structure, such as CeIn$_3$, CeTl$_3$, 
UIn$_3$, UGa$_3$, UTl$_3$ and UPb$_3$, turn antiferromagnetic 
at low temperatures. High-purity samples of CeIn$_3$, 
UIn$_3$, and UGa$_3$ have made it possible to 
characterize magnetic order and electron properties 
of these materials rather comprehensively. 
Some of the basic properties of the samples are shown 
in Table \ref{table:UCeM3}. 

\begin{figure}[h]
\centerline{
   \mbox{\includegraphics[width=1.7in]{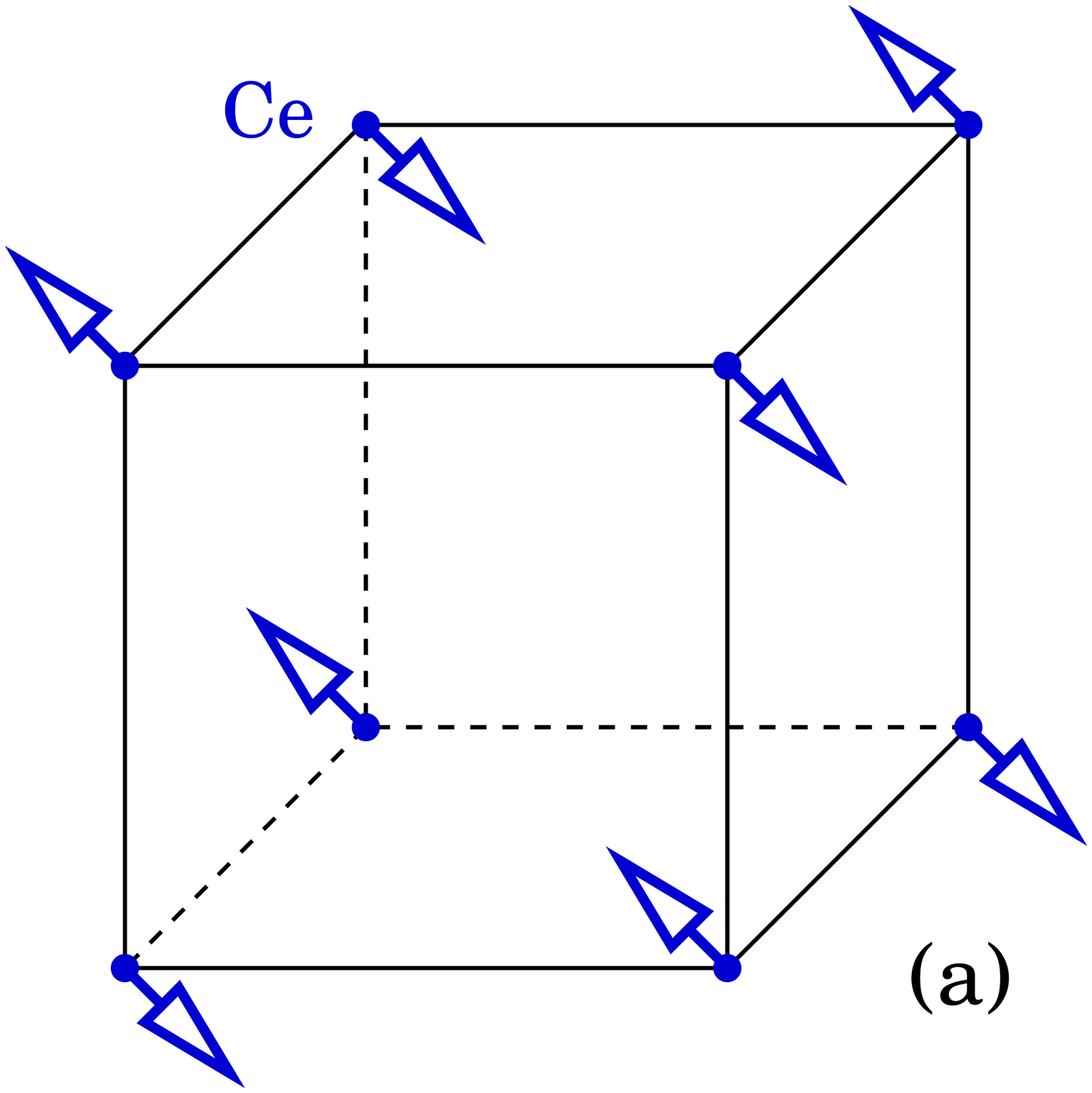}}
 \hspace{0cm}
   \mbox{\includegraphics[width=1.9in]{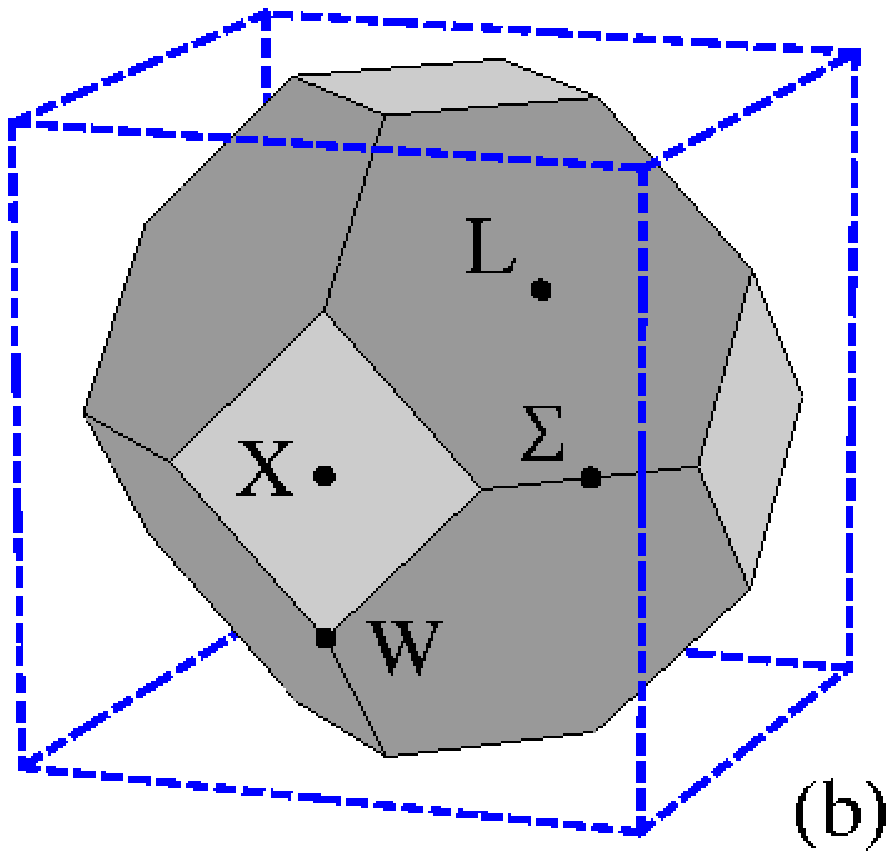}}
 \hspace{-0cm}
 }
 \caption{(color online). 
Geometry of CeIn$_3$ in real and 
in reciprocal space.  
(a) Cubic unit cell of CeIn$_3$, showing Ce atoms and 
their magnetic moments. Indium atoms (not shown) are 
positioned at the face centers of the unit cell. 
(b) Cubic Brillouin zone of paramagnetic CeIn$_3$ 
and, inside, its antiferromagnetic counterpart. 
Darker shading marks the degeneracy surface. 
}
 \label{fig:CeIn3}
\end{figure}

At low temperature, all three develop a type-II 
antiferromagnetic structure with wave vector 
${\bf Q} = \left[\frac{1}{2} \frac{1}{2} \frac{1}{2} \right]$, 
shown in Fig. \ref{fig:CeIn3}(a) for CeIn$_3$. The materials 
remain normal metals down to the lowest temperatures probed,  
with the Sommerfeld coefficient substantially enhanced by  
comparison with that of a simple metal (see the fourth 
co\-lumn of Table \ref{table:UCeM3} versus about 0.65  
mJ/K$^2\cdot$mol for Ag).

\begin{table}
\begin{center}
\caption{
Simple properties of some of the studied samples 
of CeIn$_3$, UGa$_3$ and UIn$_3$: the N\'eel 
temperature $T_N$, the ordered magnetic moment $M$, 
the Sommerfeld coefficient $\gamma$, the residual 
resistivity $\rho_0$, and the residual resistivity 
ratio $\rho(300K)/\rho_0$. 
}
\begin{tabular}{c|c|c|c|c|c}
\hline
\hline
   &  & & & & \\
  Material & $T_N$ & M & $\gamma$ & 
  $\rho_0$ &
  $\frac{\rho(300K)}{\rho_0}$  \\
   & (K) & ($\mu_B$) & (mJ/K$^2 \cdot$mol) & 
  ($\mu \Omega$cm) & \\
   &  & & & & \\
\hline
 CeIn$_3$  & 10.1 & 0.5/Ce & 130 &
 0.5 \cite{knebel} &  35 \cite{knebel}  \\
   &  & \cite{lawrence2} & \cite{knebel} &
  0.6 \cite{ebihara1} & 100 \cite{grosche1} \\
   &  & & & & \\
 UGa$_3$  &  64 & 0.75/U & 52 &
  1.2 \cite{aoki} & 38 \cite{cornelius}  \\
   & \cite{dervenagas} & \cite{dervenagas} & 
\cite{maaren,cornelius,aoki} & & 81 \cite{aoki} \\
   &  & & & & \\
 UIn$_3$  & 88 & 1/U \cite{murasik} &  50 \cite{maaren} & 
 0.66 \cite{tokiwa} & 130 \cite{tokiwa}  \\
 & & & & &  \\
%%%%%%%%%%%%%%%
\hline
\hline
\end{tabular}
\label{table:UCeM3}
\end{center}
\end{table} 

%%%%%%% Figure %%%%%%%
%%%%%%%%%%%%%%%%%%%%%% 

Of the three materials, CeIn$_3$ has been scrutinized 
the most. Its early studies were driven by interest in 
valence \cite{lawrence1} and magnetic \cite{lawrence2}  fluctuations, in the nature of its magnetic order  
\cite{lawrence2}, in large mass enhancement \cite{nasu} 
and related questions. Subsequent research focused on the 
reduction of $T_N$ under pressure, and on superconductivity, 
discovered near the critical pressure $p_c$, where the N\'eel 
temperature is about to vanish -- as well as on marked 
departure from Landau Fermi liquid behavior, found in 
the normal state near $p_c$ \cite{walker,mathur,grosche1}. 
The most recent work included de Haas-van Alphen oscillation 
measurements \cite{ebihara1,endo}, electron-positron 
annihilation experiments \cite{biasini}, and 
interpretation of the former \cite{gorkov}. 

According to Fig. \ref{fig:CeIn3}(b), the Magnetic Brillouin 
Zone of the three metals enjoys full cubic symmetry. Its 
square faces belong to the paramagnetic Brillouin zone 
boundary $({\bf p \cdot a})=\pm \pi$, where Eqn. 
(\ref{eq:thetaTU}) does not enforce degeneracy; however,  
$g_\perp ({\bf p})$ does vanish at the hexagonal MBZ 
faces, marked by darker shading in Fig. \ref{fig:CeIn3}(b). 

According to de Haas-van Alphen measurements 
\cite{ebihara1,ebihara2} and to calculations \cite{rusz}, 
one sheet of the Fermi surface of CeIn$_3$ is 
nearly spherical, and has radius of about  
$\frac{\pi}{a}\frac{\sqrt3}{2}$, where $a$ 
is the lattice constant. Hence this sheet comes 
close to the point L in Fig. \ref{fig:CeIn3}(b), 
which is the very same distance  
$\frac{\pi}{a}\frac{\sqrt3}{2}$ away from the 
Brillouin zone center. Disappearance of $g_\perp ({\bf p})$ 
necessarily affects the dynamics of an electron on this 
sheet in a transverse field.

Near a generic point at an MBZ face, far from its edges,  
leading terms of the expansion of $g_\perp ({\bf p})$ are 
linear in transverse deviation of momentum from the MBZ 
face as per Eqn. (\ref{eq:expand_g}), with ${\bf \Xi_p}$ 
normal to the MBZ boundary. Near the edges, joining the 
neighboring hexagonal faces in Fig. \ref{fig:CeIn3}(b) 
 -- for instance, near the points $\Sigma$ and $W$
 -- the leading terms become quadratic. 

\subsection{Uranium nitride} 

\begin{figure}[h]
\centerline{
   \mbox{\includegraphics[width=1.6in]{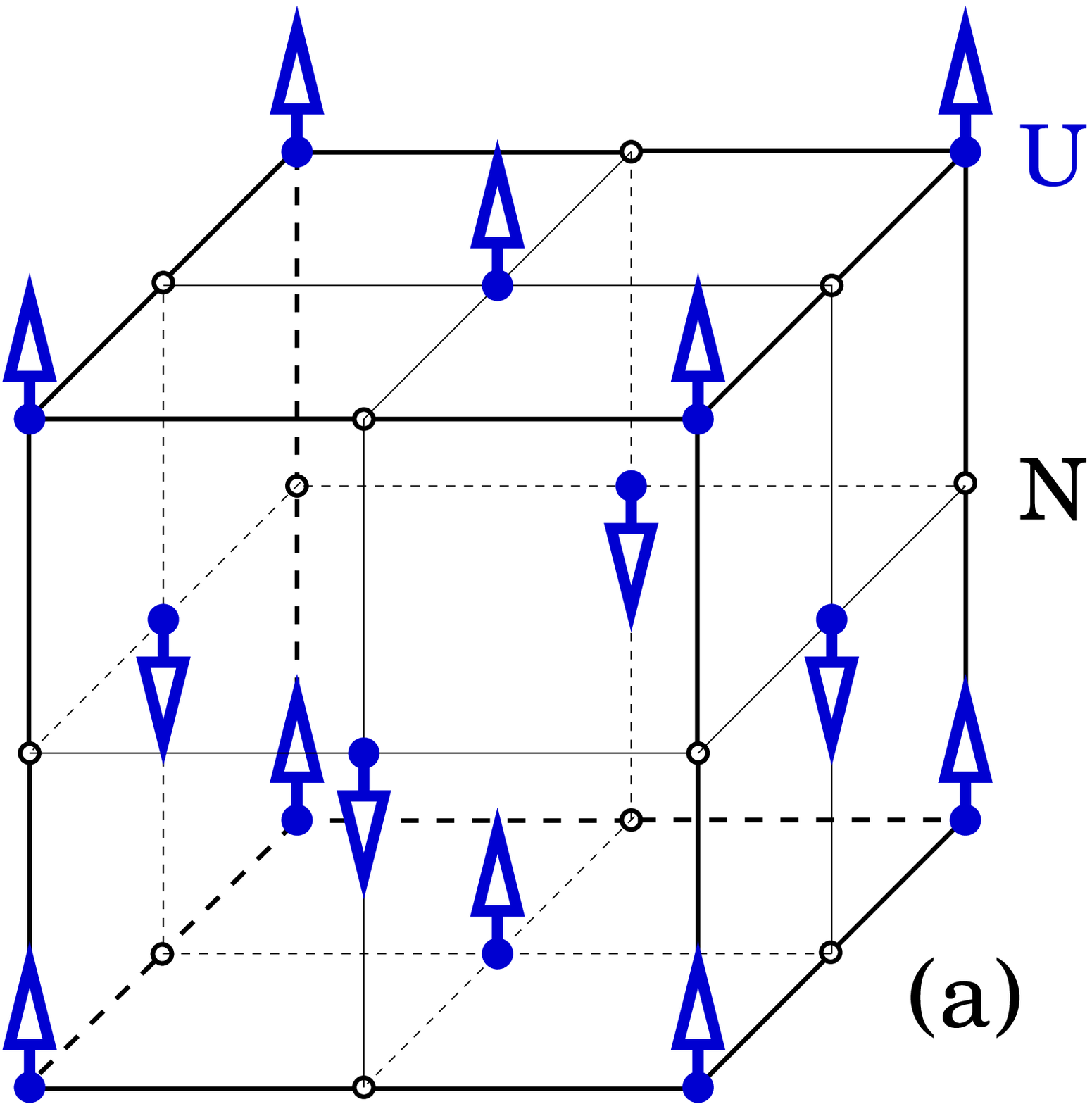}}
 \hspace{-0.5cm}
   \mbox{\includegraphics[width=1.9in]{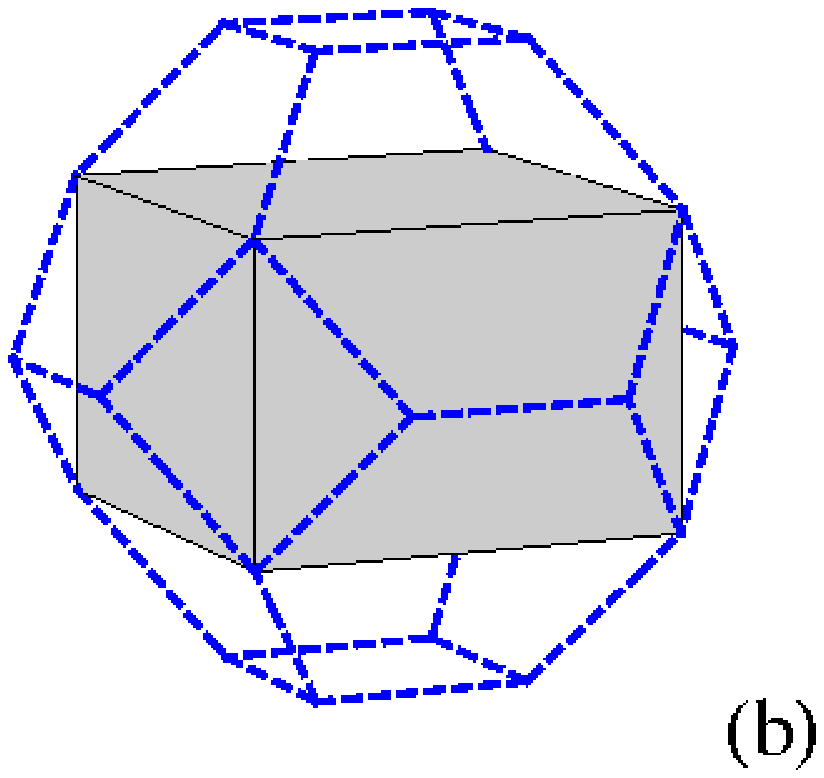}}
 \hspace{-1.5cm}
 }
 \caption{(color online). 
Geometry of uranium nitride (UN) in real 
and in reciprocal space. 
(a) F.c.c. cubic unit cell of UN, showing U 
atoms and their magnetic moment orientation. 
Nitrogen atoms are shown by open circles. 
(b) Dashed lines define the Brillouin zone boundary 
of paramagnetic UN; the square prism inside it is 
the antiferromagnetic Brillouin zone. Its entire 
boundary defines the degeneracy surface 
$g_\perp({\bf p}) = 0$.
}
 \label{fig:UN}
\end{figure}

Uranium nitride (UN) presents another example of interest. 
This  heavy fermion metal has a face-centered cubic lattice 
of NaCl type, shown in Fig. \ref{fig:UN}(a). Below 53K, 
it develops type-I antiferromagnetic order, with ordered 
moment of about 0.75$\mu_B$ per uranium atom \cite{curry}, 
and the Sommerfeld coefficient of 50 mJ/K$^2 \cdot$mol  
\cite{nakashima2}. The Ne\'eel temperature of UN drops 
under pressure, vanishing at about 3.5 GPa. 
Recent experiments \cite{nakashima2} studied the 
low-temperature resistivity near the critical pressure 
on samples with residual resistivity $\rho_0$ of about 
2.3 $\mu \Omega$cm, and the residual resistivity ratio 
$\rho(300K)/\rho_0$ of the order of $10^2$. 

%%%%%%% Figure %%%%%%%
%%%%%%%%%%%%%%%%%%%%%%

The real-space sketch of magnetic structure of UN 
is shown in Fig. \ref{fig:UN} together with its 
paramagnetic and antiferromagnetic Brillouin zone 
boundaries. The MBZ has full tetragonal symmetry  
and, in a transverse field, all the states at its 
boundary retain Kramers degeneracy. The leading terms 
in the expansion of $g_\perp({\bf p})$ are linear near 
the MBZ faces, quadratic near the edges, and cubic 
near the vertices.

\subsection{CePd$_2$Si$_2$ and CeRh$_2$Si$_2$} 

\begin{figure}[h]
\centerline{
   \mbox{\includegraphics[width=1.3in]{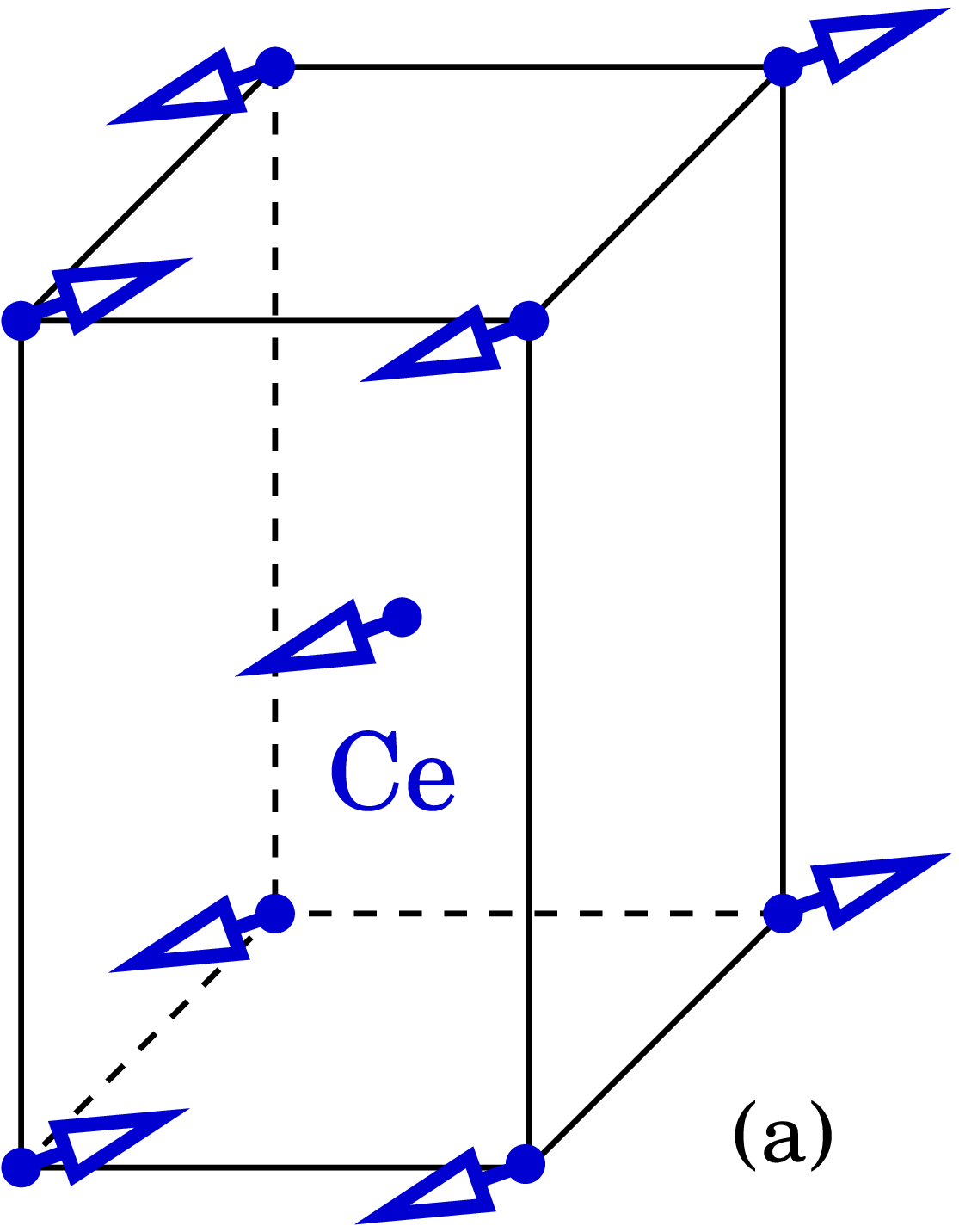}}
 \hspace{0.5cm}
   \mbox{\includegraphics[width=1.6in]{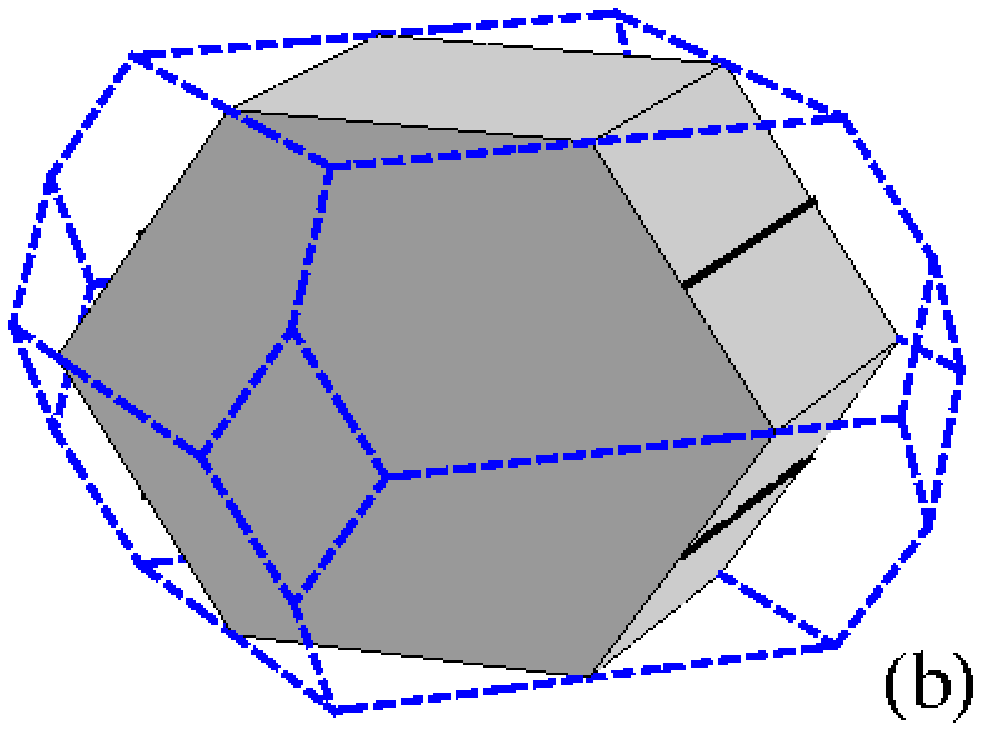}}
 \hspace{-0.5cm}
 }
 \caption{(color online). Geometry of CePd$_2$Si$_2$ 
in real and in reciprocal space. 
(a) Tetragonal unit cell of CePd$_2$Si$_2$, 
showing Ce atoms and the orientation of their 
magnetic moments. A full sketch, showing Pd 
and Si atoms, is given in \cite{grier}. 
(b) The Brillouin zone boundary of paramagnetic 
CePd$_2$Si$_2$ is shown by dashed lines. The 
shaded hexagonal prism is its antiferromagnetic 
counterpart. 
}
 \label{fig:CePd2Si2}
\end{figure}

The heavy fermion metal CePd$_2$Si$_2$ has a body-centered 
tetragonal structure of ThCr$_2$Si$_2$ type, shown in Fig. 
\ref{fig:CePd2Si2}(a). It is isostructural to CeCu$_2$Si$_2$ 
 -- the first discovered heavy fermion superconductor 
\cite{steglich} -- and CeCu$_2$Ge$_2$, an incommensurate  
antiferromagnet \cite{knopp}, that becomes superconducting 
above 70 kbar in a pressure cell \cite{jaccard}. 

Below about 10 K, CePd$_2$Si$_2$ orders antiferromagnetically 
as shown in Fig. \ref{fig:CePd2Si2}(a), with wave vector 
${\bf Q} = \left[ \frac{1}{2} \frac{1}{2} 0 \right]$, 
and a low-temperature ordered moment of about 0.7 $\mu_B$ 
per Ce atom. Its Sommerfeld coefficient is enhanced 
to about 100 mJ/K$^2 \cdot$mol.  
Samples of the present generation show residual 
resistivity in the  $\mu \Omega \cdot$cm range 
\cite{grosche1}. Under hydrostatic pressure of 
26 kbar, the N\'eel temperature drops to under 
1 K and, in a pressure window of $\pm$5 kbar 
around this value, superconductivity appears, 
with a maximum transition temperature of about 
0.4 K \cite{grosche}. Curiously enough, normal state 
resistivity near this pressure follows a temperature dependence, that does not fit the 
$\rho (T) = \rho_0 + A T^2$ temperature dependence 
of the Landau Fermi liquid theory, but instead 
behaves as $\rho(T) \sim T^{1.2}$ over more than a decade  
in temperature, between about 1 and 40 K \cite{grosche}. 

%%%%%%% Figure %%%%%%%
%%%%%%%%%%%%%%%%%%%%%%

The unit cell of CePd$_2$Si$_2$, and its paramagnetic 
and antiferromagnetic Brillouin zone boundaries are 
shown in Fig. \ref{fig:CePd2Si2}. By symmetry, the 
degeneracy manifold in a transverse field includes 
the two hexagonal faces of the MBZ boundary, 
one of which is shown by darker shading in Fig. 
\ref{fig:CePd2Si2}(b), and the four segments, two 
of which are shown in black. Along these segments, 
which are a three-dimensional analogue of point 
$\Sigma$ in Fig. \ref{fig:BZ_rectangular}(a), 
another sheet of the degeneracy surface crosses 
the side faces of the MBZ.   
According to de Haas-van Alphen (dHvA) experiments 
\cite{sheikin}, several Fermi surface sheets cross 
the degeneracy surface. The leading term in the 
expansion of $g_\perp({\bf p})$  around the hexagonal 
MBZ faces is linear. 

CeRh$_2$Si$_2$ is an isostructural relative of 
CePd$_2$Si$_2$, with a modestly enhanced Sommerfeld 
coefficient of about 23 mJ/K$^2 \cdot$mol. Between 
$T_{N1} \approx 36$K and $T_{N2} \approx 25$K, 
it develops N\'eel order with 
${\bf Q} = \left[ \frac{1}{2} \frac{1}{2} 0 \right]$ 
\cite{grier,kawarazaki}. Magnetic structure below 
$T_{N2}$ has not yet been established unambiguously 
\cite{grier,kawarazaki}. Both $T_{N1}$ and $T_{N2}$ drop 
under pressure \cite{ohashi} and, in an extended pressure 
window above 5 kbar, CeRh$_2$Si$_2$ becomes superconducting 
at a $T_c$ with a maximum of about 0.5K \cite{movshovich}.  
Antiferromagnetic structure of CeRh$_2$Si$_2$ between $T_{N1}$ 
and $T_{N2}$ coincides with that of CePd$_2$Si$_2$, as does 
the degeneracy surface in Fig. \ref{fig:CePd2Si2}(b).  
According to \cite{araki}, at least one sheet of the Fermi 
surface of CeRh$_2$Si$_2$ crosses the degeneracy surface 
or comes close to it.

\subsection{Neodymium hexaboride}

Rare earth hexaborides RB$_6$ are an interesting fa\-mi\-ly, 
whose members show diverse electron and magnetic 
properties. Of the (relatively) simple ones, LaB$_6$ is a 
diamagnetic metal, and SmB$_6$ is a mixed valence 
semiconductor. Of the ordered materials, EuB$_6$ is a 
ferromagnetic semi-metal, and CeB$_6$ is a heavy fermion 
metal with at least two ordered phases, whose nature 
remains to be elucidated after nearly forty years of research. 

Three members of the family: NdB$_6$, GdB$_6$, and 
PrB$_6$, are antiferromagnetic at low temperature. 
In PrB$_6$ \cite{burlet} and in GdB$_6$ 
\cite{mcmorrow,galera} alike, two different 
low-temperature antiferromagnetic 
states have been found. 

%%%%%%% Figure %%%%%%%
\begin{figure}[h]
\centerline{
   \mbox{\includegraphics[width=1.3in]{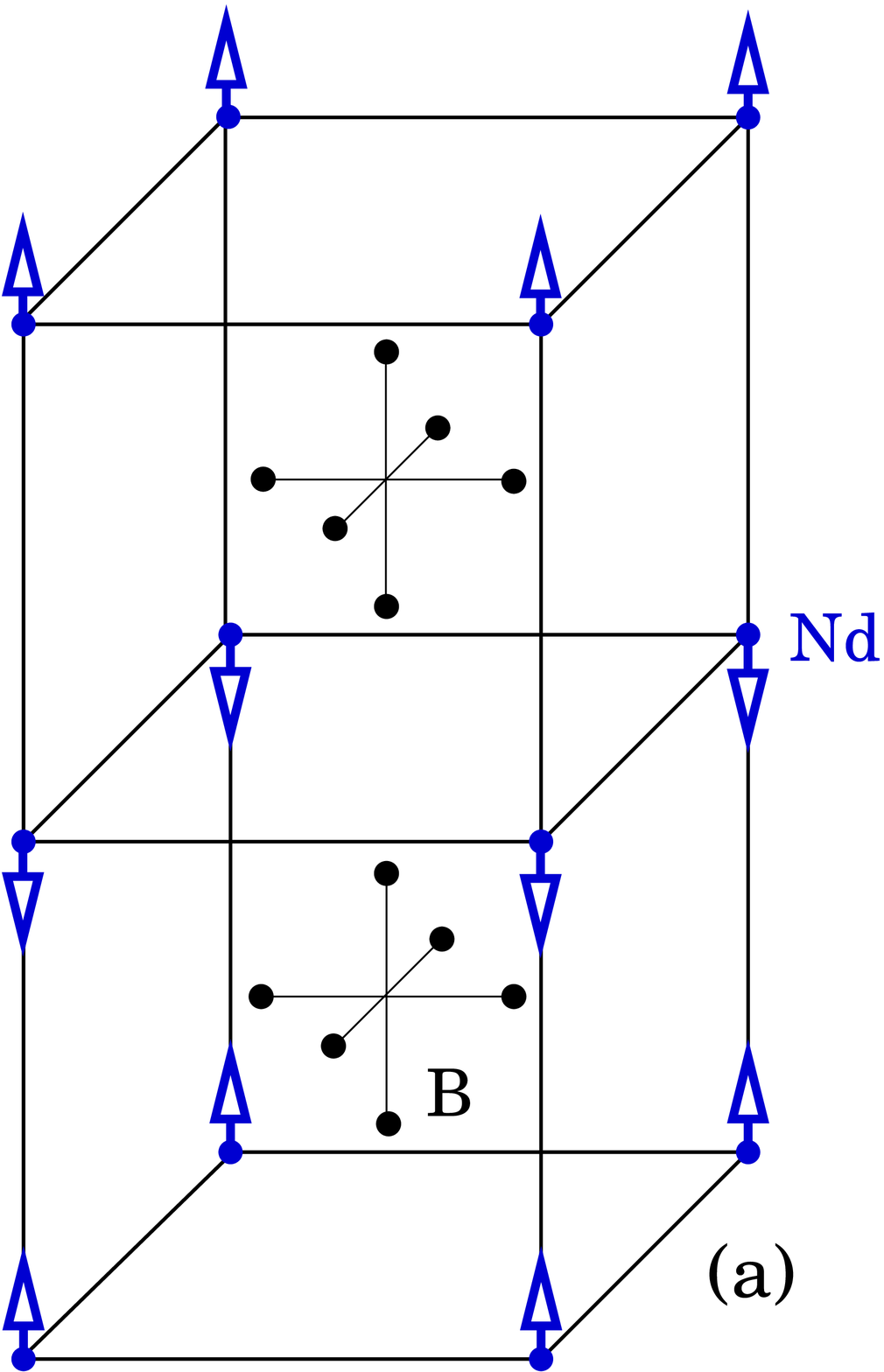}}
 \hspace{0.5cm}
   \mbox{\includegraphics[width=1.6in]{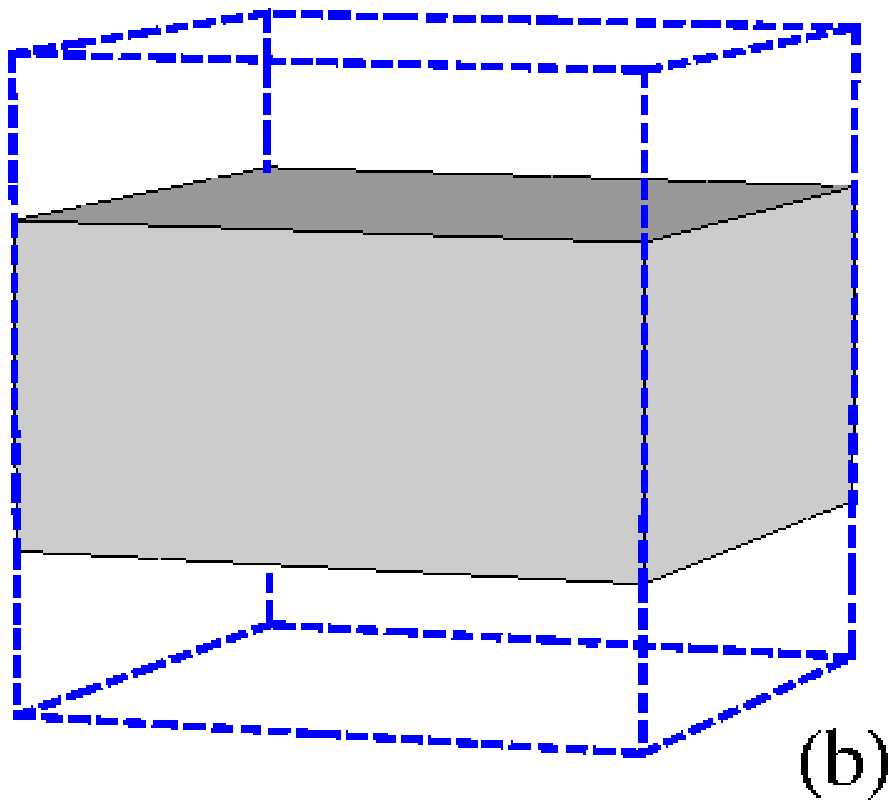}}
 \hspace{-0.5cm}
 }
 \caption{(color online). Neodymium hexaboride. 
(a) Crystalline and magnetic structure of NdB$_6$: 
a CsCl structure with B octahedra replacing Cl atoms, 
and Nd in place of Cs. The arrows show magnetic 
moments of neodymium atoms. 
(b) Dashed lines show the cubic Brillouin zone 
of paramagnetic NdB$_6$. The shaded square 
prism inside it is the tetragonal Brillouin 
zone boundary in the antiferromagnetic state. 
Its darker face denotes the degeneracy plane. 
}
 \label{fig:NdB6}
\end{figure}
%%%%%%%%%%%%%%%%%%%%%% 

Neodymium hexaboride NdB$_6$ presents a simpler 
picture: below about 8 K, it is a collinear 
type-I antiferromagnet with ordering vector 
${\bf Q} = \left[ 0 0 \frac{1}{2} \right]$ and an 
ordered moment of about 1.74 $\mu_B$ \cite{mccarthy};  
antiferromagnetism doubles its cubic unit cell in the 
[0 0 1] direction, as shown in Fig. \ref{fig:NdB6}(a). 
Thus the cubic magnetic Brillouin zone reduces 
by half in the [0 0 1] direction, while keeping 
its other two dimensions intact, as shown 
in Fig. \ref{fig:NdB6}(b). 

In a transverse field, the Kramers degeneracy 
is protected at the two faces of the MBZ boundary, 
one of which is shown by darker shading in 
Fig. \ref{fig:NdB6}(b). 
According to de Haas-van Alphen measurements 
\cite{onuki,goodrich} and to calculations \cite{kubo}, 
at least one sheet of the Fermi surface crosses the 
degeneracy surface. Recently studied samples had 
residual resistivities well below $\mu \Omega \cdot$cm, 
and residual resistivity ratios of over a 100  
\cite{onuki,stankiewicz}.

\subsection{Other materials of interest}

This subsection contains a brief discussion of other 
antiferromagnets, where symmetry may protect the 
degeneracy of special electron states against 
transverse magnetic field, giving rise to Zeeman 
spin-orbit coupling (\ref{eq:ZSO}). 

\textbf{Cuprate superconductors:} Electron-doped cuprates 
such as Nd$_{2-x}$Ce$_x$CuO$_{4 \pm \delta}$ develop 
commensurate antiferromagnetic order in a wide range 
of doping \cite{motoyama}, albeit with a modest 
staggered moment \cite{matsuda}. For such materials, 
Fig. \ref{fig:BZ_rectangular}(b) describes the 
paramagnetic and antiferromagnetic Brillouin zone boundaries. 
Angle-resolved photoemission experiments \cite{armitage} 
on Nd$_{2-x}$Ce$_x$CuO$_{4 \pm \delta}$ have found carriers 
in a vicinity of the MBZ boundary. In a transverse magnetic 
field, these carriers are subject to Zeeman spin-orbit 
coupling (\ref{eq:ZSO}), provided antiferromagnetism 
in the sample is developed well enough.

Recent observation \cite{doiron} of magnetic oscillations 
in YBa$_2$Cu$_3$O$_{6.5}$ testifies to great progress 
in sample quality of cuprates. And the fact that this and 
other underdoped cuprates are, at the very least, close 
to commensurate antiferromagnetism, makes them an 
interesting opportunity to examine the effects of 
Zeeman spin-orbit coupling. 

\textbf{Borocarbides} RT$_2$B$_2$C with 
R = Sc, Y, La, Th, Dy, Ho, Er, Tm or Lu 
and T = Ni, Ru, Pd or Pt have been a subject 
of active research, driven by interest in 
interplay between antiferromagnetism and 
superconductivity \cite{mueller}. At low 
temperatures, commensurate antiferromagnetism 
develops in a number of borocarbides 
(for instance, in $R$Ni$_2$B$_2$C with 
$R=$ Pr, Dy or Ho), often with a large 
staggered moment ($\approx$ 8.5 $\mu_B$ 
for Dy and Ho) \cite{mueller}. 
Zeeman spin-orbit coupling (\ref{eq:ZSO}) 
is present whenever a sheet of the Fermi 
surface crosses the degeneracy manifold, 
and successful growth of high-quality 
single crystals \cite{cava} makes these 
materials an interesting case to study. 

\textbf{Organic conductors} are an immense and ever 
growing class of quasi-low-dimensional materials, that 
show virtually all known types of electron states, 
found in condensed matter physics \cite{chaikin}. 
Antiferromagnetism appears in several families of 
organic conductors, and manifestations of Zeeman 
spin-orbit coupling (\ref{eq:ZSO}) are likely 
to be found in some of them. 

Unfortunately, so far nearly all of the 
information on magnetic structure of organic 
antiferromagnets has been coming from indirect 
probes such as magnetic sus\-cep\-ti\-bi\-li\-ty 
measurements \cite{iwasa,uozaki} and resonance 
spectroscopies \cite{coulon,miyagawa,wzietek}. 
Neutron diffraction studies are hampered by a typically 
small ordered moment, and by the difficulties of growing 
large enough single-crystalline samples. At the moment 
of writing, I am aware of only a single cycle of neutron 
scattering experiments \cite{foury,pouget,pouget2} 
on an organic conductor. Moreover, in families 
such as (TMTSF)$_2$ Bechgaard salts \cite{chaikin} 
and $\kappa$-(BEDT-TTF)$_2$X salts \cite{lefebvre,kagawa},  
antiferromagnetic states are insulating, and their
controlled doping remains a challenge \cite{kanoda}. 

With this word of caution, a number of organic conductors 
may deserve attention. Semi-metallic Bechgaard salt 
(TMTSF)$_2$NO$_3$ \cite{biskup}, developing a spin 
density wave state below about 9 K, may be one interesting 
case. Recently synthesized  
ethylenedioxytetrathiafulvalenoquinone-1,3-diselenolemethide 
(EDO-TTFVODS), that appears to turn antiferromagnetic below 
about 4.5 K, and remains normal down to the lowest studied 
temperature of 0.45 K \cite{xiao}, may be another. 
Finally, recent studies \cite{suzuki,zhou,hara} 
of [Au(tmdt)$_2$], where tmdt denotes trimethylenetetrathiafulvalenedithiolate, draw 
attention to this organic conductor. Albeit the material 
is not yet fully characterized, and its large single 
crystals remain difficult to grow, it appears to have 
a N\'eel temperature of about 110 K \cite{zhou,hara}, 
which is anomalously high for an organic material
 -- and shows normal conduction down to at least 10 K.

\textbf{Heavy fermion materials:} Several heavy 
fermion antiferromagnets were reviewed in detail 
above. A number of other interesting examples 
may be found in \cite{robinson}. 

\textbf{Gadolinium antiferromagnets} 
(see \cite{barandiaran,granado}, and Table I 
in \cite{rotter}) offer two important advantages 
for an experimental study of the Zeeman spin-orbit 
coupling. Firstly, their often elevated N\'eel 
temperature $T_N$ (such as 134 K for GdAg, or 150 K 
for GdCu) facilitates experimental access to 
temperatures well below $T_N$, where thermal 
fluctuations of antiferromagnetic order are frozen 
out. Secondly, large ordered moment of these materials 
(about 7.5$\mu_B$ for GdAg, and about 7.2$\mu_B$ 
for GdCu$_2$Si$_2$) quenches quantum fluctuations. 
Therefore, gadolinium antiferromagnets fit well 
into the present framework with its neglect 
of both quantum and classical fluctuations 
 -- and shall be convenient for a study of 
various effects of the Zeeman spin-orbit coupling. 

\textbf{Iron pnictides} have been attracting immense 
attention \cite{grant} due to appearance of commensurate 
antiferromagnetism \cite{cruz} and high-temperature 
superconductivity \cite{takahashi} in this copper-free 
family of materials. Combination of commensurate 
antiferromagnetism \cite{cruz} with essentially metallic 
normal state conduction \cite{takahashi,kamihara} 
not only contrasts iron pnictides with the cuprates 
(that are believed to be Mott insulators), but also 
makes the former materials likely to manifest a 
substantial momentum dependence of the $g$-tensor.

\section{V. Discussion} 
\subsection{Effects of relativistic spin-orbit coupling} 

The arguments above appealed to the exchange symmetry 
approximation \cite{andreev}: the point symmetry 
operations of the electron Hamiltonian in an 
antiferromagnet were considered inert with respect 
to spin, and the common  relativistic spin-orbit 
coupling, that appears in the absence of an external 
magnetic field, was thus neglected. 
I will now examine the effects it may have.  
 
Firstly, this spin-orbit interaction generates magnetic 
anisotropy, that creates a preferential orientation 
of the staggered magnetization ${\bf n}$ with respect 
to the crystal axes. In an experiment, this allows 
one to vary the magnetic field orientation with 
respect to ${\bf n}$ as long as the field remains 
below the reorientation threshold.

At the same time, the spin-orbit coupling may eliminate 
those spatial symmetries, that rotate the magnetization 
density with respect to the lattice. For instance, 
certain spin rotations and spatial transformations, 
that were independent symmetries within the exchange 
symmetry approximation, may survive only when combined. 
I will now illustrate this by two examples of Section IV. 

A simple case of the spin-orbit coupling affecting 
the Kramers degeneracy manifold in a transverse field 
is given by a two-dimensional antiferromagnet 
on a square-symmetry lattice as in Fig.  
\ref{fig:BZ_rectangular}(b). Here, the Kramers 
de\-ge\-ne\-ra\-cy at the antiferromagnetic 
Brillouin zone boundary relies, everywhere 
except for points $\Sigma$, on the symmetry 
with respect to reflections in diagonal planes 1 and 2. 
% diagonal reflections $\sigma_{1,2}$  
If either of these reflections changes the orientation  
of $\bm\Delta_{\bf r}$ with respect to the crystal 
axes, a spin-orbit coupling may lift the degeneracy 
at a relevant part of the magnetic Brillouin zone 
boundary, except for points $\Sigma$. 
However, if the magnetization density points along 
one of these diagonal axes, the degeneracy survives 
at the two faces of the MBZ boundary, that are 
normal to this axis.

In the case of commensurate order in chromium, 
consider a single-domain sample with magnetic 
structure shown in Fig. \ref{fig:Cr_real_space}. 
For a Bloch state $| {\bf p} \rangle$ with a momentum 
${\bf p}$ at one of the two horizontal faces of the 
magnetic Brillouin zone in Fig \ref{fig:Cr}(a), the 
degenerate partner state 
$\theta {T_a U_n}(\pi) | {\bf p} \rangle$  
has momentum $- {\bf p}$ at the other horizontal 
face of the MBZ. Coordinate rotation by $\pi$ around 
the vertical symmetry axis, passing through the center 
$\Gamma$ of the Brillouin zone, transforms the momentum 
${\bf p}$ into ${\bf p + Q}$, equivalent to ${\bf p}$ 
up to reciprocal lattice vector ${\bf Q}$ 
of the antiferromagnetic state. 

By contrast, for a momentum ${\bf p}$ at one of the 
vertical faces of the MBZ, coordinate rotation by $\pi$ 
around a horizontal axis is required, and such a rotation 
inverts ${\bf \Delta_r}$ once the latter is attached to 
the crystal axes. Thus spin-orbit coupling tends to lift 
the degeneracy at the vertical faces of the MBZ, leaving 
it intact at the two horizontal faces. The other examples 
of Section III can be analyzed similarly. 

Finally, those spin-orbit coupling terms, that 
act directly on the electron spin and tend to lift 
the double degeneracy of Bloch eigenstates even 
in the absence of magnetic field, were neglected 
here altogether.

\subsection{Relation to earlier work}  
When symmetries of a system involve time reversal 
 -- alone or in combination with other operations -- 
a proper treatment must involve non-unitary symmetry 
groups: those containing unitary as well as anti-unitary 
elements. In this case, construction of irreducible 
representations is complicated by the fact that 
anti-unitary elements involve complex conjugation. 
In a group representation, combination of two unitary 
elements $u_1$ and $u_2$ is represented by 
the pro\-duct of the corresponding matrices 
${\bf D}(u_1)$ and ${\bf D}(u_2)$: 
${\bf D}(u_1 u_2) = {\bf D}(u_1) {\bf D}(u_2)$. 
By contrast, combination of an anti-unitary element 
$a$ with a unitary element $u$ involves complex 
conjugation: ${\bf D}(a u) = {\bf D}(a) {\bf D}^*(u)$. 
As a result, irreducible representations of a non-unitary 
group must include a unitary representation and its 
complex conjugate on an equal footing. Discussion of 
such representations (called co-representations) was 
given by Wigner, along with the analysis of arising 
possibilities with the help of the Frobenius-Schur 
criterion \cite{wigner}. Later, Herring studied 
spectral degeneracies, emerging in crystals due to time 
reversal symmetry and, among other things, extended this 
criterion to space groups \cite{herring2}. In a subsequent 
work, Dimmock and Wheeler generalized the criterion further, 
to magnetic crystals, and pointed out the sufficient 
condition (\ref{eq:Dimmock_equivalence}) for the 
appearance of extra degeneracies \cite{dimmock2}. 
 
The present work identifies the symmetry, that protects  
the Kramers degeneracy in a N\'eel antiferromagnet 
against transverse magnetic field, as a conspiracy 
between the anti-unitary symmetry 
${\bf U_n}(\pi) {\bf T_a} \theta$, inherent to any 
collinear commensurate antiferromagnet in a transverse 
field, and the crystal symmetry of those special 
momenta at the MBZ boundary, that are defined by 
Eqn. (\ref{eq:Dimmock_equivalence}).  
Formally, the present work is an extension of 
\cite{dimmock2}, since one may think of the last 
two terms in (\ref{eq:Hamiltonian_generic}) as of 
the exchange field of a fictitious magnetic crystal 
in zero field. 
However, Kramers degeneracy in a magnetic field 
has rather special and remarkable experimental 
signatures, some of which are outlined at the 
end of this section. 

% Last but not the least, article \cite{braluk}
% by Brazovskii and Lukyanchuk was of utmost 
% importance for the present work. Its authors
% studied electron eigenstates in a N\'eel 
% antiferromagnet on a square lattice, built the 
% co-representations of interest, and analyzed their 
% properties in the vicinity of various high-symmetry
% points in the Brillouin zone. The authors also studied
% the lifting of the Kramers degeneracy by magnetic field 
% and, among other things, pointed out the disappearance
% of $g_\perp$ at the MBZ boundary, as well as the ensuing 
% substantial momentum dependence of $g_\perp$ in the 
% Zeeman coupling (\ref{eq:ZSO}). Unfortunately, the
% analysis \cite{braluk} is not entirely satisfactory; 
% the details can be found in Appendix D. 
 
Last but not the least, Ref. \cite{braluk} 
was an important source of inspiration for 
the present work. Its authors studied the 
electron eigenstates in a N\'eel antiferromagnet 
on a lattice of square symmetry and, for this 
par\-ti\-cu\-lar case, pointed out the 
disappearance of $g_\perp({\bf p})$ at the MBZ 
boundary, as well as the ensuing substantial momentum 
dependence of $g_\perp$ in the Zeeman coupling 
(\ref{eq:ZSO}). The present article builds on 
Ref. \cite{braluk} by elucidating the structure 
of the manifold of degenerate states for an 
arbitrary crystal symmetry, and for an arbitrary 
transverse field that can be sustained by the 
antiferromagnet before its sublattices collapse. 
This is to be contrasted with the analysis of 
Ref. \cite{braluk}, performed to the linear 
order in the field. Several other aspects of 
Ref. \cite{braluk} are discussed in Appendix D. 

\subsection{Experimental signatures} 
The Kramers degeneracy at special momenta on the 
MBZ boundary and the resultant Zeeman spin-orbit 
coupling have a number of interesting consequences. 
For instance, a substantial momentum dependence 
of $g_\perp ({\bf p})$ in Eqn. (\ref{eq:ZSO}) 
means that, generally, the Electron Spin 
Re\-so\-nance (ESR) frequency of a carrier in the 
vicinity of the de\-ge\-ne\-ra\-cy manifold varies 
along the quasiclassical trajectory in momentum space. 

For a weakly-doped antiferromagnetic insulator with 
a conduction band minimum on the degeneracy manifold, 
this leads to an inherent broadening of the ESR line 
with do\-ping and, eventually, complete loss 
of the ESR signal. In fact, this may well be 
the reason behind the long-known `ESR silence'  \cite{shengelaya} of the cuprates. Suppression 
of Pauli paramagnetism in the transverse direction 
with respect to staggered magnetization is another 
simple consequence of vanishing $g_\perp({\bf p})$. 

At the same time, a momentum dependence of 
$g_\perp({\bf p})$ allows excitation of spin 
resonance transitions by AC {\em electric} rather 
than magnetic field \cite{rr,zedr} -- a vivid effect 
of great promise for controlled spin manipulation, 
currently much sought after in spin electronics. 
Its absorption matrix elements are defined by 
${\bf \Xi}_{\bf p}$ of Eqn.  (\ref{eq:expand_g}). 
Comparison with Eqn. (\ref{eq:weak_coupling_g}) 
shows that, within the weak-coupling model  
(\ref{eq:weak_coupling_Hamiltonian}), 
$\Xi_{\bf p}/\hbar$ is of the order 
of the antiferromagnetic coherence length 
$\xi \sim \frac{\hbar v_F}{\Delta}$, and 
may be of the order of the lattice period 
or much greater. By contrast, the ESR matrix 
elements are defined by the Compton length 
$\lambda_C = \frac{\hbar}{mc} \approx 0.4$ pm. 
Thus, matrix elements of electrically excited 
spin transitions exceed those of ESR by about 
$\frac{\hbar c}{e^2} \cdot \frac{\epsilon_F}{\Delta} 
\approx 137 \cdot \frac{\epsilon_F}{\Delta}$, 
or at least by two orders of magnitude. Being 
proportional to the square of the appropriate 
transition matrix element, resonance absorption 
due to electric excitation of spin transitions 
exceeds that of ESR at least by four orders of 
magnitude.

Last but not the least -- according to 
Eqn. (\ref{eq:expand_g}), resonance absorption 
in this phenomenon shows a non-trivial dependence 
on the orientation of the AC electric field with 
respect to the crystal axes, and on the orientation 
of the DC magnetic field with respect to the 
staggered magnetization. 

The Zeeman spin-orbit coupling may also manifest itself 
in other experiments on antiferromagnetic conductors. 
In particular, de Haas-van Alphen oscillations 
\cite{kabanov} and magneto-optical response may 
be modified. In va\-ri\-ous types of electron response, 
interesting effects may arise due to an extra term 
${\bf v}_{ZSO}$ in the electron ve\-lo\-ci\-ty 
ope\-ra\-tor, emerging due to a substantial 
momentum dependence of $g_\perp({\bf p})$ 
in Eqn. (\ref{eq:ZSO}): 
\be
\label{eq:velocity_bit} 
{\bf v}_{ZSO} = \nabla_{\bf p} \mathcal{H}_{ZSO} 
 = - \mu_B \nabla_{\bf p} g_\perp ({\bf p}) 
({\bf H_\perp \cdot \bm{\sigma}}). 
\ee
This term describes spin current. However, 
$g_\perp ({\bf p})$ is even in ${\bf p}$ 
due to inversion symmetry and thus, in 
equilibrium, the net spin current must 
vanish. This may change, if the system 
were tilted, say, by electric current 
or otherwise -- however, the resulting 
effect would be proportional to the `tilt' 
and, in addition to this, would be small 
in the measure of $H_\perp/\Delta$.

\subsection{Conclusions}

In this work, I studied the degeneracy of electron 
Bloch states in a N\'eel antiferromagnet, subject 
to a transverse magnetic field -- and described 
the special points in momentum space, where the 
degeneracy is protected by a hidden anti-unitary 
symmetry. 

I discussed the simplest properties and some of the 
manifestations of the Zeeman spin-orbit coupling, 
arising in a magnetic field due to this degeneracy, 
and outlined se\-ve\-ral examples of interesting 
materials, where such a coupling may be present.  
Finally, I reviewed the results and their 
relation to earlier work. 

The degeneracy of special Bloch states in a transverse field 
hinges only on the symmetry of the antiferromagnetic state, 
and thus holds in weakly coupled and strongly correlated 
materials alike -- provided long-range antiferromagnetic 
order and well-defined electron quasiparticles are present. 
Under these conditions, thermal and quantum fluctuations 
of the antiferromagnetic order primarily renormalize the 
sublattice magnetization, leaving intact the degeneracy 
of special electron states in a transverse field
 -- certainly in the leading order in fluctuations. 
Detailed account of fluctuations is outside 
the scope of this article.

\section{Acknowledgments}

I am indebted to S. Brazovskii and G. Shlyapnikov 
for inviting me to Orsay, to LPTMS for the kind 
hos\-pi\-ta\-li\-ty, and to IFRAF for generous 
support. I am grateful to S. Brazovskii, 
M. Kartsovnik and K. Kanoda for re\-fe\-ren\-ces 
on organic antiferromagnets, and to C. Capan for 
drawing my attention to gadolinium compounds. 
It is my pleasure to thank S. Carr, N. Cooper, 
N. Shannon, and M. Zhi\-to\-mir\-sky for their 
helpful comments on the manuscript, 
and A. Chu\-bu\-kov and G. Volovik 
for enlightening discussions.  

\section{Appendix A: orthogonality relation}

This Appendix proves the relation 
\be
\label{eq:antiproduct}
\langle
\phi |
\left[ \mathcal{O} \theta \right]^+ 
     | 
 \left[
 \mathcal{O} \theta 
 \right]
 | 
\psi
\rangle
 = 
\langle 
\psi 
 | 
\phi
\rangle, 
\ee 
where $| \phi \rangle$ and $| \psi \rangle$ 
are arbitrary states,  $\mathcal{O}$ 
is an arbitrary unitary operator, and $\theta$ 
is time reversal. In the main text, this relation 
is used for 
$| \phi \rangle = \mathcal{O} \theta | \psi \rangle$; 
in this case, when read right to left, 
Eqn. (\ref{eq:antiproduct}) yields 
\be
\label{eq:antiproduct2}
\langle 
\psi 
 |
\mathcal{O} \theta | \psi \rangle 
 =
\langle
\psi 
 |
[ \left(\mathcal{O} \theta \right)^+ ]^2
 | 
 (
 \mathcal{O} \theta )
 |
\psi
\rangle.
\ee  
Whenever $| \psi \rangle$ is an 
eigenvector of the linear operator 
$[ \mathcal{O} \theta ]^2$ with an 
eigenvalue different from unity, 
Eqn. (\ref{eq:antiproduct2}) proves 
orthogonality of $| \psi \rangle$ and 
$ \mathcal{O} \theta | \psi \rangle$. 

The proof of Eqn. (\ref{eq:antiproduct}) is based 
on the obvious relation 
$(\mathcal{C} \phi, \mathcal{C} \psi)
 = (\psi, \phi)$ for arbitrary complex vectors 
$\phi$ and $\psi$, where 
$(\psi, \phi) \equiv \sum_i \psi_i^* \phi_i$ 
denotes scalar product, and $\mathcal{C}$ 
is complex conjugation. 
Hence, for an arbitrary unitary operator 
$\mathcal{O}$, one finds 
$(\mathcal{O} \mathcal{C} \phi, \mathcal{O} \mathcal{C} \psi) 
 = (\psi, \phi)$, due to invariance of scalar product 
under unitary transformation. Time reversal $\theta$ 
can be presented as a product of $\mathcal{C}$ and a 
unitary operator \cite{wigner}: 
$\theta = \mathcal{V} \mathcal{C}$ 
, thus 
$\mathcal{C} = \mathcal{V}^{-1} \theta$ and, therefore, 
$(\mathcal{O} \theta \phi, \mathcal{O} \theta \psi) 
 = (\psi, \phi)$. 
As a result, for arbitrary states 
$| \psi \rangle$ and $| \phi \rangle$,  one finds 
$
\langle
\phi |
\left[ \mathcal{O} \theta \right]^+ 
     | 
 \left[
 \mathcal{O} \theta 
 \right]
 | 
\psi
\rangle
 = 
\langle 
\psi 
 | 
\phi
\rangle
$, 
which indeed amounts to (\ref{eq:antiproduct}).

\section{Appendix B: canting of the sublattices}

Canting of the two sublattices by transverse 
field ${\bf H}_\perp$ induces a component 
$\bm{\Delta}_{\bf r}^\perp$ 
of the magnetization density along the field, 
with the periodicity of the underlying lattice: 
$\bm{\Delta}_{\bf r+a}^\perp (H_\perp)
 = \bm{\Delta}_{\bf r}^\perp (H_\perp)$, 
as shown in Fig. \ref{fig:triad_canted}. 
As a result, the diagonal part of Hamiltonian 
(\ref{eq:weak_coupling_Hamiltonian}) acquires 
an additional term 
$(\bm{\Delta}_{\bf p}^\perp \cdot \bm{\sigma})$, 
and Hamiltonian (\ref{eq:weak_coupling_Hamiltonian}) 
thus takes the form 
\be
\label{eq:weak_coupling_Hamiltonian_canted}
\mathcal{H} 
 = 
\left[
\begin{array}{cc}
\epsilon_{\bf p}
 - ({\bf \tilde\Delta_{p}}^\perp \cdot \bm{\sigma})
 & 
({\bf \Delta}_\| \cdot \bm{\sigma}) \\
 & \\
({\bf \Delta}_\| \cdot \bm{\sigma}) 
 & 
\epsilon_{\bf p + Q}
 - ({\bf \tilde\Delta_{p+Q}}^\perp \cdot \bm{\sigma})
\end{array}
\right],
\ee
where ${\bf \tilde\Delta_{p}}^\perp \equiv 
\bf{H}_\perp + \bm{\Delta}_{\bf p}^\perp$. 

The same choice of spin axes as in Section III splits 
Hamiltonian (\ref{eq:weak_coupling_Hamiltonian_canted}) 
into two independent pieces 
\be
\label{eq:canted_H12}
\mathcal{H}_{1(2)} 
 = 
\left[
\begin{array}{cc}
\epsilon_{\bf p} \mp \tilde\Delta_{\bf p}^\perp
 &    \Delta_\| \\
 & \\
      \Delta_\| 
 & 
\epsilon_{\bf p + Q} \pm 
\tilde\Delta_{\bf p+Q}^\perp 
\end{array}
\right]. 
\ee
As in Section III, momentum shift by ${\bf Q}$
maps $\mathcal{H}_1$ and $\mathcal{H}_2$ onto 
each other, and spectral symmetries of Hamiltonian 
(\ref{eq:weak_coupling_Hamiltonian_canted}) 
coincide with those discussed in the second 
subsection of Section III. Thus all of the 
conclusions of Section III remain valid after 
sublattice canting is accounted for. 

However, while degeneracy at special points is 
protected by symmetry, the shape of the manifold 
of degenerate states may change under various 
perturbations. For instance, sublattice canting 
in a transverse field mo\-di\-fies the equation, 
describing this manifold and, for the conduction 
band, turns it into 
\be 
\label{eq:degeneracy_manifold_canted}
\psi_{\bf p} + 
\frac{\phi_{\bf p}\zeta_{\bf p}}{\sqrt{\Delta^2_\|+\phi_{\bf p}^2+\zeta_{\bf p}^2}},
\ee
where $\phi_{\bf p} \equiv \frac{1}{2} 
[ \tilde\Delta_{\bf p}^\perp + 
\tilde\Delta_{\bf  p+Q}^\perp ]$ 
and 
$\psi_{\bf p} \equiv \frac{1}{2} 
[ \tilde\Delta_{\bf p}^\perp -  
\tilde\Delta_{\bf  p+Q}^\perp ]$. 
Since $\bm{\Delta}_{\bf r}^\perp$ 
has the real-space periodicity of the paramagnetic 
state, $\tilde\Delta_{\bf p}^\perp$ enjoys the same 
reciprocal space symmetry as $\epsilon_{\bf p}$. In  
particular, $\tilde\Delta_{{\bf p}+2{\bf Q}}^\perp
 = \tilde\Delta_{\bf p}^\perp$, and 
$\tilde\Delta_{\bf p}^\perp = 
 \tilde\Delta_{-\bf p}^\perp$ (the latter property 
is also protected by the ${\bf U_l}\theta$ symmetry). 
At the same time, $\psi_{\bf p+Q} = - \psi_{\bf p}$,  
and $\phi_{\bf p+Q} = \phi_{\bf p}$; thus 
the symmetry-dictated degeneracy points such 
as ${\bf p} = \frac{\bf Q}{2}$ explicitly 
belong to the manifold of Eqn.  
(\ref{eq:degeneracy_manifold_canted}), as they should.

In the limit of vanishing $H_\perp$, 
$\tilde\Delta_{\bf p}^\perp$ is linear in the field: 
$\tilde\Delta_{\bf p}^\perp = H_\perp [1 + \chi_{\bf p}^\perp]$, 
where $\chi_{\bf p}^\perp$ describes microscopic 
transverse susceptibility of the antiferromagnet.  
Now one may expand Eqn. (\ref{eq:degeneracy_manifold_canted}) 
to linear order in the field to obtain the 
following equation for the degeneracy manifold:
\be
\label{eq:degeneracy_manifold_canted-linearized} 
\chi_{\bf p}^- + 
\frac{\chi_{\bf p}^+ \zeta_{\bf p}}{\sqrt{\Delta^2_\|
 + \zeta_{\bf p}^2}} = 0,
\ee
where 
$\chi_{\bf p}^\pm \equiv \chi_{\bf p} \pm \chi_{\bf p+Q}$. 
Compared with equation $\zeta_{\bf p} = 0$ of Section III,  
the sublattice canting affects the de\-ge\-ne\-ra\-cy 
manifold already in zeroeth order in $H_\perp$. 

\section{Appendix C: Dimensionality of the degeneracy manifold}

The dimensionality of the degeneracy manifold in 
a transverse field is one less than that of the 
momentum space for simple reasons, that rely 
only on the symmetry of the antiferromagnetic 
state. According to Eqn. 
(\ref{eq:Kramers}), zero-field Bloch eigenstates 
$| 1 \rangle \equiv | {\bf p} \rangle$ and 
$| 2 \rangle \equiv \mathcal{I}{\bf T_a}
 \theta | {\bf p} \rangle$ 
form a Kramers doublet at momentum ${\bf p}$. 
Its splitting $\delta\mathcal{E}({\bf p})$ in 
a transverse field ${\bf H}_\perp$ is given by 
\be 
\label{eq:splitting_generic} 
\delta\mathcal{E}({\bf p})
 = 2 \sqrt{|V_{12}({\bf p})|^2 + 
\frac{1}{4}
\left[V_{11}({\bf p}) - V_{22}({\bf p}) \right]^2}, 
\ee
where $V_{ij}({\bf p}) \equiv 
\langle i|({\bf H}_\perp \cdot {\bm\sigma})|j \rangle$. 
Magnetic field being uniform, 
$({\bf H}_\perp \cdot {\bm\sigma})$ commutes 
with $\mathcal{I}{\bf T_a}$; it also changes 
sign under time reversal. 
Thus, $V_{22}({\bf p}) = - V_{11}({\bf p})$. 
At the same time, the off-diagonal matrix element 
$V_{12}({\bf p})$ va\-ni\-shes identically: 
\bea 
\nonumber
\langle {\bf p} | 
({\bf H}_\perp \cdot {\bm\sigma}) 
\mathcal{I}{\bf T_a} \theta 
 | {\bf p} \rangle
   = 
% \nonumber
% & = & 
\sum_{\bf q} 
\langle {\bf p} | 
({\bf H}_\perp \cdot {\bm\sigma}) 
| {\bf q} \rangle \langle {\bf q} |
\mathcal{I}{\bf T_a} \theta 
 | {\bf p} \rangle 
 & = & 
   \\
% \nonumber
\label{eq:off-diagonal} 
 = 
\sum_{\bf q} 
V_{11}({\bf p}) \delta_{\bf p q} 
 \langle {\bf q} |
\mathcal{I}{\bf T_a} \theta 
 | {\bf p} \rangle 
   =
%   \\
% & = & 
V_{11}({\bf p}) 
 \langle {\bf p} |
\mathcal{I}{\bf T_a} \theta 
 | {\bf p} \rangle 
 \equiv  0,
\eea
where insertion of unity ${\bf 1} = \sum_{\bf q} 
| {\bf q} \rangle \langle {\bf q} |$ was used in 
the first line, uniformity of ${\bf H}_\perp$ 
in the second, and the final equality followed   
from Eqn. (\ref{eq:Kramers}). Therefore, 
\be
\label{eq:splitting}
\delta\mathcal{E}({\bf p}) = 2 |V_{11}({\bf p})|,  
\ee 
and, barring a special case, equation  
$\delta\mathcal{E}({\bf p})=0$ defines a 
$(d-1)$-dimensional surface of zero $g_\perp({\bf p})$ 
in $d$-dimensional momentum space. The Kramers 
degeneracy subset contains, at the very least, 
the star of the momentum ${\bf p} = {\bf Q}/2$ 
[see Eqn.(\ref{eq:Dimmock_equivalence}) and the 
subsequent discussion], and the $({\bf k \cdot p})$ 
expansion \cite{kittel} around these points shows, 
that they are not isolated, but rather belong 
to a $(d-1)$-dimensional manifold. The latter 
is continuous, with the obvious exception of $d = 1$.

Finally, notice that, according to (\ref{eq:splitting}), 
$\delta\mathcal{E}({\bf p})$ is periodic with the  
antiferromagnetic ordering wave vector ${\bf Q}$: 
\be
\delta\mathcal{E}({\bf p}+{\bf Q})
 = \delta\mathcal{E}({\bf p}), 
\ee 
thanks to ${\bf Q}$ being a reciprocal lattice 
vector in the antiferromagnetic state. Therefore, 
properties (\ref{eq:Q-boost_for_g}) and 
(\ref{eq:around_Q/2}) are indeed model-independent, 
as opposed to hinging on an approximation of the 
weak-coupling model (\ref{eq:weak_coupling_Hamiltonian}).

\section{Appendix D: revisiting \cite{braluk}}

In Ref. \cite{braluk}, Brazovskii and Lukyanchuk 
stated that %, in the absence of magnetic field, 
operator ${\bf \Lambda} = ({\bf n} \cdot \bm{\sigma})$ 
exchanges the momenta ${\bf p}$ and ${\bf p + Q}$ 
in (\ref{eq:weak_coupling_Hamiltonian}), and 
thus represents the momentum boost by the 
ordering wave vector ${\bf Q}$ in reciprocal space. 
In a commensurate antiferromagnetic state, ${\bf Q}$ 
becomes a reciprocal lattice vector, and thus 
${\bf \Lambda}$ must be a symmetry of the Hamiltonian. 
With the assumption of the effective Zeeman coupling 
(\ref{eq:ZSO}), this lead the authors of Ref. 
\cite{braluk} to the relation 
\mbox{$g_\perp({\bf p + Q}) = - g_\perp({\bf p})$} 
(Eqn. (\ref{eq:Q-boost_for_g}) of the present work), 
and to the conclusion that $g_\perp({\bf p}) = 0$ 
at the MBZ boundary. Unfortunately, while this 
beautiful result is indeed correct for a lattice 
of square symmetry, several circumstances 
prevent one from embracing these arguments. 
% of Ref. \cite{braluk}. 

Most importantly, they hinge solely on the 
symmetry under translation by ${\bf Q}$, 
put other\-wise -- on commen\-su\-ra\-bi\-li\-ty 
of magnetic order with the crystal lattice. 
If correct, this would imply that, in an 
arbitrary commensurate N\'eel antiferromagnet, 
Kramers degeneracy takes place at the entire 
MBZ boundary regardless of the underlying 
crystal symmetry. The toy example 
(\ref{eq:hopping}) with $\eta \neq 1$ shows, 
that this is not at all necessarily the case. 

Indeed, for a generic crystal symmetry, the condition 
\mbox{$g_\perp({\bf p + Q}) = - g_\perp({\bf p})$} 
(see \cite{braluk} and Eqn. (\ref{eq:Q-boost_for_g}) 
of the present work) does \textit{not}, by itself, 
restrict the ma\-ni\-fold $g_\perp({\bf p}) = 0$ 
to the MBZ boundary. However, the `Kramers' subset 
of the manifold of degenerate states can be obtained 
by combining Eqn. (\ref{eq:Q-boost_for_g}) with the 
crystal symmetries (see the subsection on spectral 
symmetries in momentum space in Section III). 
For instance, combined with the inversion symmetry 
\mbox{$g_\perp(- {\bf p}) = g_\perp({\bf p})$}, 
Eqn. (\ref{eq:Q-boost_for_g}) stipulates that 
\mbox{$g_\perp({\bf Q}/2) = 0$}. Similarly,  
disappearance of $g_\perp({\bf p})$ at the 
entire MBZ boundary for the square symmetry 
case can be obtained by using 
Eqn. (\ref{eq:Q-boost_for_g}) and the 
point symmetries of the square lattice. 
For a finite as opposed to infinitesimal field, 
these results were established in Section II 
and in the first two examples in Section IV. 

On a more technical level, the operator ${\bf \Lambda}$ 
is equivalent to ${\bf U_n}(\pi)$ and thus inverts the 
sign of the transverse component of the field. Hence, 
in a field with non-zero transverse component 
${\bf H}_\perp$, ${\bf \Lambda}$ ceases to be a symmetry 
of the Hamiltonian, in agreement with the first line of 
Table \ref{table:symmetries} -- and thus can no longer 
represent the momentum boost by ${\bf Q}$.  

\begin{details}
In spite of this, the authors of Ref. \cite{braluk} 
obtained the correct relation 
\mbox{$g_\perp({\bf p + Q}) = - g_\perp({\bf p})$} 
(Eqn. (\ref{eq:Q-boost_for_g}) of the present work): 
while alluding to the non-existent symmetry operator 
${\bf \Lambda}$, in fact they used essentially the 
Eqn. (\ref{eq:my_symmetry}), even if without derivation.

Th beautiful result is indeed correct for a lattice 
of square symmetry, but only due to the conspiracy 
between Eqn. (\ref{eq:Q-boost_for_g}) and the square 
symmetry of the lattice. 

As we saw above, this beautiful result 
is indeed correct for a lattice of square 
symmetry. Unfortunately, several circumstances 
prevent one from embracing this argument. 
Most importantly, it hinges solely on the 
symmetry under translation by ${\bf Q}$, 
put other\-wise -- on commensurability of 
magnetic order with the crystal lattice. 
If correct, this argument would imply that, in 
an arbitrary commensurate N\'eel antiferromagnet, 
Kramers degeneracy takes place at the entire 
MBZ boundary regardless of the underlying 
crystal symmetry. The toy example 
(\ref{eq:hopping}) with $\eta \neq 1$ shows, 
that this is not at all necessarily the case: 
the condition 
$g_\perp({\bf p + Q}) = - g_\perp({\bf p})$ 
does \textit{not}, by itself, restrict the 
manifold $g_\perp({\bf p}) = 0$ to the MBZ 
boundary (see the subsection on spectral 
symmetries in momentum space). 

Technically, the error may be 
traced back to the following.  
Firstly, the way it is written in \cite{braluk}, 
operator ${\bf \Lambda}$ does not represent  
translation by wave vector ${\bf Q}$. 
Rather, it is 
\be
\label{eq:Sigma}
{\bf \Sigma} 
 = 
\left[
\begin{array}{cc}
 0 
 & 
({\bf n} \cdot \bm{\sigma}) \\
 & \\
({\bf n} \cdot \bm{\sigma}) 
 & 
 0
\end{array}
\right]
\ee
that, in a purely longitudinal magnetic field, 
exchanges the diagonal matrix elements of Hamiltonian 
(\ref{eq:weak_coupling_Hamiltonian}), and thus is 
equivalent to translation by ${\bf Q}$ in reciprocal space. 

Secondly, ${\bf \Sigma}$ changes the sign of 
the transverse component of the field, since 
${\bf (n \cdot \bm{\sigma}) (H \cdot \bm{\sigma}) 
(n \cdot \bm{\sigma}) = (H_\| \cdot \bm{\sigma})
 - (H_\perp \cdot \bm{\sigma})}$. 
Hence, in a field with non-zero transverse component 
${\bf H}_\perp$, ${\bf \Sigma}$ ceases to be a symmetry of 
the Hamiltonian, in agreement with the first line of Table 
\ref{table:symmetries} -- and thus no longer represents 
translation by ${\bf Q}$.
\end{details}

\end{document}